\pdfoutput=1
\documentclass{elsart}

\usepackage{cite}
\usepackage{amsmath}
\usepackage{graphicx}
\usepackage{float}

\begin{document}

\runauthor{W. Rolke}
\begin{frontmatter}
\title{A Comparison of Limit Setting Methods for the On-Off Problem}
\author[wolf]{Wolfgang A. Rolke}
\address[wolf]{Department of Mathematics, University of Puerto Rico - Mayag\"{u}ez, Mayag\"{u}ez, PR 00681, USA, 
\newline Postal Address: PO Box 3486, Mayag\"{u}ez, PR 00681, 
\newline Tel: (787) 255-1793, Email: wolfgang.rolke@upr.edu}

\begin{abstract}
   We study the frequentist properties of confidence intervals found with various methods previously proposed
   for the On-Off problem. We derive explicit formulas for the limits and calculate the true coverage and the 
   expected lengths of these methods.
\end{abstract}
\end{frontmatter}\newpage

\section{Introduction}

In this paper we study a problem called \textquotedblleft On-Off \textquotedblright\ in
astrophysics and the problem of \textquotedblleft one signal band and one sideband \textquotedblright\ in high
energy physics (HEP). In the astrophysics version, one points the telescope at
( \textquotedblleft On \textquotedblright\ ) a potential point signal source and observes $x$ counts in a
particular amount of observing time. Then one points the telescope away from
( \textquotedblleft Off \textquotedblright\ ) the source to a nearby region thought to have no point sources, and
observes $y$ counts in an observing time that is $\tau $ times as long as
the \textquotedblleft On \textquotedblright\ observing time. The latter provides an estimate of the
non-point-source rate $b$, which crudely speaking allows for a \textquotedblleft background
subtraction \textquotedblright\  in the on-source data. One desires a confidence interval for
the point source rate $\mu $ in the presence of the background rate that can
be inferred (with some uncertainty) only from $y$. Complications arise
because, for weak sources and small counts $x$ and $y$, the
Poisson-distributed data is not suitable for trivial formulas coming from
Gaussian assumptions. So the probability model we are studying is given by%
\begin{equation}
X\sim Pois(\mu +b)\text{ \ \ }Y\sim Pois(\tau b)  \label{onoff}
\end{equation}%
The same probability model arises if the background rate was estimated via
Monte Carlo, in which case $\tau $ is related to the number of Monte Carlo
runs.

Statistical Inference for this problem has a long history in HEP and
Astronomy. It received renewed interest in 1997 by Feldman and
Cousins \cite{Feldman:1997qc}, who applied their now famous Unified method to the case of 
a known background rate $b$. A
general method for including an uncertainty into the model was proposed by
Cousins and Highland \cite{Cousins-Highland} and was applied to the \textquotedblleft On-Off \textquotedblright\  in Conrad
et.al. \cite{Conrad:2002kn} and Tegenfeldt and Conrad \cite%
{Tegenfeldt:2004dk}. 

A solution based on inverting the likelihood ratio test was proposed by Rolke and Lopez \cite%
{Rolke:2000ij} and extended to include uncertainty in the efficiency as well
as other probability models for $Y$ by Rolke, Lopez and Conrad \cite%
{Rolke:2004mj}. This is also known as profile
likelihood method, which has a long history in Statistics and has been in use for a
some time in physics in the MINUIT program when errors are calculated with
the MINOS method, see James and Roos \cite{James:1975dr} and
Li and Ma \cite{Li:1983fv}. The class TRolke implements this method in ROOT,
see Lundberg et.al. \cite{Lundberg:2009iu}.

A different approach known as the CLs method was described by Read \cite%
{Read:2002hq}, and Gan and Kass \cite{Gan:1997rg} used the Cousins-Highland
prescription to include uncertainties in the background rate into CLs. This solution
is different from the others discussed in this paper in that it calculates
upper limits only. Finally a number of intervals derived via the Bayesian
paradigm have been proposed as well.

The problem of significance testing for the signal rate was studied by
Cousins, Linnemann and Tucker \cite{Cousins:2008zz}, and in principle
hypothesis testing and interval estimation are the same problem, as one can
always invert a hypothesis test into an interval procedure and vice versa.
In practice this can be difficult, and in fact the method favored by
Cousins, Linneman and Tucker, namely turning the problem into one for a
Binomial parameter p, can not be inverted into a limit setting method
because it is specific to the test $H_{0}:\mu =0$. What is missing from the
literature is a detailed study of the commonly used methods for limit
setting in terms of their coverage as well as other properties, and it is
this study we are undertaking in this paper.

The software tools commonly used to calculate these limits, for example
RooStats, are general purpose tools that allow for much more complicated
probability models, for example multiple channels, uncertainties in the
detection efficiencies etc. Such generality comes at a price as these tools
use Monte Carlo simulation for the limit calculations. This leads to two
problems for our study: on the one hand the limits come with errors, often
on the order of 5-10\%, and it is not clear how one would determine the
correct coverage in this case. Moreover, these tools are fairly slow in
calculating the limits. This is not an issue if just one or even a few
limits are needed. We, though, need a very large number of them: our study
focuses on the small sample case, and so we restrict ourselves to $\mu \in
\lbrack 0,20]$ and $b\in \lbrack 0,10]$. We also want to study a number of
different $\tau $ values, say $\tau \in \{0.5, 1,2\}$. With these parameters
possible observations for which we need the limits range from $x=0$ to $x=50$
and the same for $y$. Also we want to study coverage at the nominal $68\%$, $%
90\%$ and $95\%$ levels. This means we need to calculate a total of $23409$ limits for each method. Obviously this requires a
very fast way to find the limits, and in the next section we are
developing exact formulas to be able to do so. Unfortunately this task
becomes hopeless if we tried to incorporate further uncertainties, for
example in $\tau $, into the models.

It should be noted that all results in this paper are based on exact
calculations. They therefore do not carry with them any
uncertainties due to a finite number of simulation runs.

\section{The Methods}

\subsection{Intervals based on Inverting a Likelihood Ratio Test: Rolke-Lopez-Conrad (RLC)}

Maybe the most widespread technique for deriving a hypothesis test in
Statistics is the likelihood ratio test (LRT). Say we have a probability
model $f(\mathbf{x};\theta )$\ and we wish to test $H_{0}:$ $\theta \in
\Theta _{0}$ vs $H_{a}:$ $\theta \notin \Theta _{0}$. Then the LRT is a test
based on the test statistic%
\begin{equation}
T(\mathbf{x)=}\frac{\sup_{\Theta _{0}}f(\mathbf{x};\theta )}{\sup_{\Theta }f(%
\mathbf{x};\theta )}
\end{equation}%
where $\Theta _{0}$ is some subset of the parameter space $\Theta $.

The reason for the popularity of this approach is two-fold: first by the famous 
Neyman-Pearson lemma it is known that in the case of a simple vs.
simple hypothesis this test is optimal, that is has the highest possible
power for a given type I error probability $\alpha $. Even though optimality is not guaranteed in more complicated cases, experience has shown that tests (and intervals) derived with this method tend to do very well. Also, under some regularity conditions in the large sample limit $-2\log T(%
\mathbf{x)}$ has a chi-square distribution. This is known as Wilk's theorem,
and a proof can be found in many Statistics text books, for example in Casella and Berger \cite{casella2002statistical}.

For our probability model we have 
\begin{equation}
f(x,y;\mu ,b)=\frac{(\mu +b)^{x}}{x!}e^{-(\mu +b)}\frac{(\tau b)^{y}}{y!}%
e^{-\tau b}  \label{pdf}
\end{equation}%
so $\theta =(\mu ,b)$. Here and in what follows $\tau $ is considered a
known constant. We want a confidence interval for $\mu $ alone, so we test $%
H_{0}:$ $\mu =\mu _{0}$ vs $H_{a}:\mu \neq \mu _{0}$. For the denominator of
\thinspace $T$ \ we need to find the maximum likelihood estimators (mle),
which are $\widehat{\mu }=x-y/\tau $ and $\widehat{b}=y/\tau $. For the
numerator we need to maximize $f(x,y;\mu ,b)$ for $b$ alone, treating $\mu $
as fixed. This leads to%
\begin{equation}
\widehat{\widehat{b}}=\frac{x+y-(1+\tau )\mu +\sqrt{(x+y-(1+\tau )\mu
)^{2}+4(1+\tau )y\mu }}{2(1+\tau )}  \label{bhatmu}
\end{equation}

and then the test can be based on 
\begin{equation}
-2\log T(x,y\mathbf{)=}2\left( \log f(x,y;\widehat{\mu },\widehat{b})-\log
f(x,y;\mu _{0},\widehat{\widehat{b}})\right)
\end{equation}

Replacing a nuisance parameter by the value that maximizes the likelihood
function while keeping the parameter of interest fixed is known as the
profile likelihood method and has a long history in Statistics.

Unfortunately in our case, at least when $\mu _{0}=0$, the regularity
conditions of Wilk's theorem are not satisfied. Moreover, we are far from a
large-sample regime, and so the question arises as to what the null
distribution might be. This was studied by Rolke and Lopez \cite{Rolke:2000ij} who
showed that the chi-square approximation is surprisingly good, and that
confidence intervals derived by inverting the likelihood ratio test and using the chi-square approximation have good coverage properties, at least when some ad-hoc adjustments are made in the cases where the observed
number of events in the signal region is less than what is expected from
background alone. The $RLC$ method is implemented in the Root class Trolke,
described in Lundberg et.al. \cite{Lundberg:2009iu}.

\subsection{Feldman-Cousins Unified Method}

In 1997 Feldman and Cousins \cite{Feldman:1997qc} proposed a limit setting
method for the model $X\sim Pois(\mu +b)$ where $b$ is assumed known. The
method proceeds as follows. Consider the Poisson density%
\begin{equation}
f(x;\mu )=\frac{\mu ^{x}}{x!}e^{-\mu }
\end{equation}%
where $x=0,1,..$ and $\mu \geq 0$. Now for all possible observations $x$ and
a signal rate $\mu $ calculate the ratio%
\begin{equation}
\frac{f(x;\mu +b)}{f(x;\widehat{\mu }+b)}
\end{equation}%
where $\widehat{\mu }=\max \{0,x-b\}$ is the maximum likelihood estimator\
(mle) of $\mu $. Rank the possible observations $x$ according to these
ratios, resulting in the sequence $x_{i}^{\ast },i=1,2..$\ Define the
 \textquotedblleft acceptance region \textquotedblright\ $A(\mu )$ by%
\begin{equation}
A(\mu )=\left\{ (x_{1}^{\ast },..,x_{n}^{\ast });\sum_{i=1}^{n}f(x_{i}^{\ast
};\mu +b)\leq cl,\sum_{i=1}^{n+1}f(x_{i}^{\ast };\mu +b)>cl\right\} 
\end{equation}%
where $cl$ is the desired confidence level, for example $0.95$ for a  $95\%$ confidence interval. In the
Statistics literature this methodology for deriving a test (and a corresponding limits method) is called the Neyman
construction. The confidence interval for $\mu $ is comprised of all
values of $\mu $ such that $A(\mu )$ includes the observed $x$. Because in the problem studied here this
method leads to a simple interval this last step can be done by finding the
endpoints. 

It should be noted that in this case of a Poisson rate with a known background the limits found by the Feldman-Cousins method can also be derived by inverting the corresponding likelihood ratio test. 

How can we extend this method to our more general problem? We will consider
four options:

\subsubsection{Feldman-Cousins Confidence Regions (FCR)}

The most obvious solution is to consider confidence regions in $(\mu ,b)$
space. So now we have pairs of points $(x,y)$ and again we find their
Poisson probabilities%
\begin{equation}
f(x,y;\mu ,b)=\frac{(\mu +b)^{x}}{x!}e^{-(\mu +b)}\frac{(\tau b)^{y}}{y!}%
e^{-\tau b}
\end{equation}

We find the maximum likelihood estimators $\widehat{\mu }=\max \{0,x-y/\tau
\}$ and $\widehat{b}=y/\tau $, and calculate the ratios

\begin{equation}
\frac{f(x,y;\mu ,b)}{f(x,y;\widehat{\mu },\widehat{b})}
\end{equation}

\begin{figure}[H]
\caption{2 dimensional $95\%$ confidence region for $(\protect\mu ,b)$ if $%
x=20,y=7$ and $\protect\tau =1.0$. A $95\%$ confidence interval for $\protect%
\mu $ is found by projecting region onto $\protect\mu $ axis}
\label{fig:FCReg}
\centering %
\includegraphics[width=0.98\textwidth,natwidth=450,natheight=450]{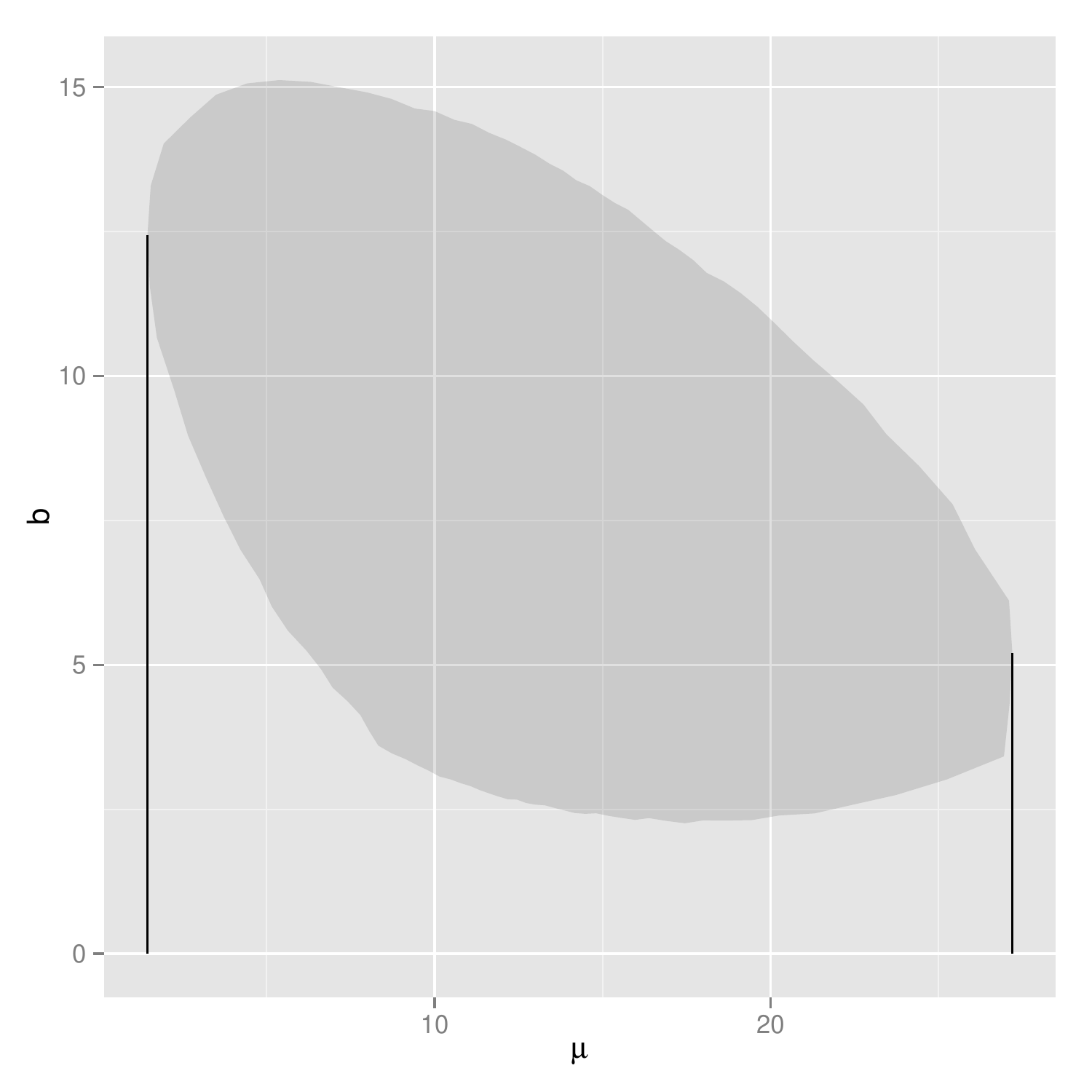}
\end{figure}

Then we rank the points according to this ratio and \textquotedblleft accept \textquotedblright\ all points up
to the desired confidence level. Then we scan through $(\mu ,b)$ space to
find the confidence region. As an example consider figure \ref{fig:FCReg}
which shows the $95\%$ confidence region for the case $x=20$, $y=7$ and $\tau
=1$.

Of course $b$ is a nuisance parameter, and what we really want is a
confidence interval for $\mu $. We can get that by projecting the region
down onto the $\mu $-axis, but it is clear that such a method will suffer
from over-coverage. Unfortunately no general method is known to extract a
confidence interval from a confidence region in such a way that the
confidence interval has correct coverage.

\subsubsection{Feldman-Cousins Profile Likelihood (FCPL)}

Because we are only interested in $\mu $ we can try to eliminate the
background rate $b$ at some point during the calculations. One way to do
this is to use the idea of profile likelihood already described above. Here
when calculating the probabilities we replace the density $f(x,y;\mu ,b)$ by 
$f_{PL}(x,y;\mu )$ defined by $f_{PL}(x,y;\mu )=f(x,y;\mu ,\widehat{\widehat{%
b}})$, where $\widehat{\widehat{b}}$ is as in equation \ref{bhatmu}. Now the
method proceeds exactly as the basic Feldman-Cousins method described above,
except using $f_{PL}$ instead of $f$.

One problem with this approach is that $f_{PL}$ is no longer a probability
density, in fact $\sum_{x,y}f_{PL}(x,y;\mu )=\infty $. So we need to
restrict our calculations to $0\leq x\leq M_{x}$ and $0\leq y\leq M_{y}$
with $M_{x}$ and $M_{y}$ large enough so that their choice does not effect
the limits significantly. In the numerical studies shown below we always use $%
M_{x}=M_{y}=50$, which we verified is large enough so that the effect on the
limits is negligible. Moreover we need to normalize $f_{PL\text{ }}$so it is
a proper probability density with%
\begin{equation}
\sum_{x=0}^{M_{x}}\sum_{y=0}^{M_{y}}f_{PL}(x,y;\mu )=1
\end{equation}

\subsubsection{Cousins-Highland}

Another way to eliminate $b$ is to use a procedure first advocated by
Cousins and Highland \cite{Cousins-Highland}. The idea is to eliminate a
nusiance parameter by integrating it out. There are essentially two ways to
do this:

\emph{Feldman-Cousins-Cousins-Highland Option 1 (FCCH1)}

Here we replace the probability density $f(x,y;\mu ,b)$ with

\begin{equation}
\begin{tabular}{l}
$f_{CH}(x,y;\mu ):=\int_{0}^{\infty }f(x,y;\mu ,b)db=$ \\ 
$\frac{\tau ^{y}}{x!y!}e^{-\mu }\int_{0}^{\infty }(\mu
+b)^{x}b^{y}e^{-(1+\tau )b}db=$ \\ 
$\frac{\tau ^{y}}{x!y!}e^{-\mu }\int_{0}^{\infty }\left( \sum_{n=0}^{x}%
\binom{x}{n}\mu ^{n}b^{x-n}\right) b^{y}e^{-(1+\tau )b}db=$ \\ 
$\frac{\tau ^{y}}{x!y!}e^{-\mu }\sum_{n=0}^{x}\binom{x}{n}\mu
^{n}\int_{0}^{\infty }b^{x+y-n}e^{-(1+\tau )b}db=$ \\ 
$\frac{\tau ^{y}}{x!y!}e^{-\mu }\sum_{n=0}^{x}\binom{x}{n}\mu ^{n}\frac{%
\Gamma (x+y-n+1)}{(1+\tau )^{x+y-n+1}}$%
\end{tabular}%
\end{equation}%
\newline
where we made use of the fact that%
\begin{equation}
\int_{0}^{\infty }t^{k}e^{-at}dt=\frac{\Gamma (k+1)}{a^{k+1}}
\end{equation}

which in turn follows because the integrand is the density of a Gamma
distribution with parameters $k+1$ and $a$. $\Gamma$ denotes the gamma
function.

Now limits are found the same way as before. Again we have the problem that $%
\sum_{x,y}f_{CH}(x,y;\mu )=\infty $, and we proceed as in section $2.2.2$.
The normalized probabilities will be denoted by $f_{CH}^{*}$.

\emph{Feldman-Cousins-Cousins-Highland Option 2 (FCCH2)}

In this version one first calculates limits $L(x;b)$ and $U(x;b)$ for the
signal rate $\mu $ using the method of Feldman and Cousins for fixed
background rate $b$, and then finds limits 
\begin{equation}
\begin{tabular}{l}
$L(x,y)=\int_{0}^{\infty }L(x;b)\tau \frac{(\tau b)^{y}}{y!}e^{-\tau b}db$
\\ 
$U(x,y)=\int_{0}^{\infty }U(x;b)\tau \frac{(\tau b)^{y}}{y!}e^{-\tau b}db$%
\end{tabular}%
\end{equation}

essentially weighting each limit by the corresponding Poisson probabilities.
The extra factor $\tau $ comes from the normalization $\int_{0}^{\infty }%
\frac{(\tau b)^{y}}{y!}e^{-\tau b}db=\frac{1}{\tau }$.

\subsubsection{Neyman Construction with Probability Ordering (NeyProb)}

There is yet another variation of this method: instead of using the
likelihood ratio as the ordering principle we can simply use the
probabilities $f_{CH}^{\ast }$. One of the reasons Feldman and Cousins did
not use this ordering was that it can lead to empty intervals. For example
if we have $\tau =1$, observe $x=0,y=6$ and want to find $95\%$ limits there
is no $\mu \geq 0$ that will have the point $(0,6)$ in the acceptance
region. If this is deemed acceptable, or if it is known a priori that there
will be more events in the signal region than are expected from background
alone, this is a viable method.

\subsection{CLs}

A method that has been used extensively in HEP is the CLs method. Here in
the case of a known background rate $b$ one uses the test statistic%
\begin{equation}
\frac{P(X\leq x|\mu +b)}{P(X\leq x|b)}
\end{equation}%
which means one is testing specifically $H_{0}:$ $\mu =0$ vs $H_{a}:\mu >0$.
Therefore this method always yields upper limits.

CLs was first proposed in a special case by Zech \cite{Zech:1988un} and generalized
by Read \cite{Read:2002hq}. Even though it has very little grounding
in Statistical theory it has become quite popular in HEP. The extension to
our \textquotedblleft On-Off \textquotedblright\ model was done by Gan and Kass \cite{Gan:1997rg}, who used the
Cousins-Highland prescription to show that an upper limit can be found by
inverting a test based on the test statistic%
\begin{equation}
T_{CLs}(\mu ;x,y,\tau )=\frac{e^{-\mu }}{y!}\frac{\sum_{n=0}^{x}%
\sum_{k=0}^{n}\mu ^{n-k}\binom{y+k}{y}/(n-k)!/(1+\tau )^{k}}{\sum_{n=0}^{x}%
\binom{y+k}{y}/(1+\tau )^{k}}  \label{cls}
\end{equation}%
and the (say) $95\%$ upper limit is found by solving $T_{CLs}(\mu ;x,y,\tau
)=0.95$. The derivation of equation \ref{cls} is very similar to the
calculations done for $f_{CH}$.

\subsubsection{Bayesian Methods}

The last class of methods we will consider are intervals derived using the Bayesian approach.
Although we will find proper Bayesian credible intervals these will then be
evaluated as standard frequentist confidence intervals. Helene \cite%
{Helene:1982pb} used a Bayesian approach to calculate limits for the \textquotedblleft On-Off \textquotedblright\
problem, although they modeled the background as a Gaussian rather than a
Poisson. 

As always in Bayesian Statistics we need to choose priors for $\mu $ and $b$. We will consider the following cases: flat priors, a fairly common choice in HEP,
and Jeffrey's non-informative prior, which in the case of a Poisson
distribution with rate $\mu $ is given by $\pi (\mu )=1/\sqrt{\mu }$. We
will consider all combinations of these priors. Lastly we will include
another choice occasionally found in the literature, namely $\pi (\mu ,b)=%
\frac{1}{\sqrt{\mu +b}}$. This is Jeffreys prior when $b$ is a known constant.

The last case does not allow for an analytic solution, and limits are found
through numerical integration. For the first four we can handle all cases in
one calculation by assuming priors of the form $\pi (\lambda )=\lambda
^{-\rho }$, where $\rho =0$ or $1/2$. \ With this we have the probability model

\begin{equation}
\begin{tabular}{l}
$X\sim Pois(\mu +b)$, $Y\sim Pois(b)$ \ \ $\pi (b)=b^{-\rho }$ $\pi (\mu
)=\mu ^{-\gamma }$ \\ 
$f(x,y;\mu ,b)=\frac{(\mu +b)^{x}}{x!}e^{-(\mu +b)}\frac{(\tau b)^{y}}{y!}%
e^{-\tau b}b^{-\rho }\mu ^{-\gamma }=$ \\ 
$\frac{\tau ^{y}}{x!y!}(\mu +b)^{x}b^{y-\rho }\mu ^{-\gamma }e^{-\mu
-(1+\tau )b}$, \ $x,y=0,1,..;\mu ,b\geq 0$%
\end{tabular}%
\end{equation}%
we begin by finding the marginal distribution of $x$ and $y$:
\begin{equation}
\begin{tabular}{l}
$m(x,y)=\int_{0}^{\infty }\int_{0}^{\infty }f(x,y;\mu ,b)\pi (b)\pi (\mu
)d\mu db=$ \\ 
$\frac{\tau ^{y}}{x!y!}\int_{0}^{\infty }b^{y-\rho }e^{-(1+\tau )b}\left(
\int_{0}^{\infty }(b+\mu )^{x}\mu ^{-\gamma }e^{-\mu }d\mu \right) db=$ \\ 
$\frac{\tau ^{y}}{x!y!}\int_{0}^{\infty }b^{y-\rho }e^{-(1+\tau )b}\left(
\int_{0}^{\infty }\left[ \sum_{n=0}^{x}\binom{x}{n}b^{n}\mu ^{x-n}\right]
\mu ^{-\gamma }e^{-\mu }d\mu \right) db=$ \\ 
$\frac{\tau ^{y}}{x!y!}\sum_{n=0}^{x}\binom{x}{n}\int_{0}^{\infty
}b^{y+n-\rho }e^{-(1+\tau )b}\left( \int_{0}^{\infty }\mu ^{x-n-\gamma
}e^{-\mu }d\mu \right) db=$ \\ 
$\frac{\tau ^{y}}{x!y!}\sum_{n=0}^{x}\binom{x}{n}\int_{0}^{\infty
}b^{y+n-\rho }e^{-(1+\tau )b}\Gamma (x-n-\gamma +1)db=$ \\ 
$\frac{\tau ^{y}}{x!y!}\sum_{n=0}^{x}\binom{x}{n}\Gamma (x-n-\gamma
+1)\int_{0}^{\infty }b^{y+n-\rho }e^{-(1+\tau )b}db=$ \\ 
$\frac{\tau ^{y}}{x!y!}\sum_{n=0}^{x}\binom{x}{n}\Gamma (x-n-\gamma +1)\frac{%
\Gamma (y+n-\rho +1)}{(1+\tau )^{y+n-\rho +1}}$%
\end{tabular}%
\end{equation}%
Next we find the posterior density of $\mu $ as the marginal of the joint posterior density:%
\begin{equation}
\begin{tabular}{l}
$f(\mu |x,y)=\int_{0}^{\infty }f(\mu ,b|x,y)db=$ \\ 
$\int_{0}^{\infty }\frac{f(x,y,;\mu ,b)}{m(x,y)}db=$ \\ 
$\int_{0}^{\infty }\frac{\tau ^{y}}{m(x,y)x!y!}(\mu +b)^{x}b^{y-\rho }\mu
^{-\gamma }e^{-\mu -(1+\tau )b}db=$ \\ 
$\frac{\tau ^{y}}{m(x,y)x!y!}\int_{0}^{\infty }(\mu +b)^{x}b^{y-\rho }\mu
^{-\gamma }e^{-\mu -(1+\tau )b}db=$ \\ 
$\frac{\tau ^{y}}{m(x,y)x!y!}e^{-\mu }\mu ^{-\gamma }\int_{0}^{\infty
}\sum_{n=0}^{x}\binom{x}{n}b^{n}\mu ^{x-n}b^{y-\rho }e^{-(1+\tau )b}db=$ \\ 
$\frac{\tau ^{y}}{m(x,y)x!y!}e^{-\mu }\sum_{n=0}^{x}\binom{x}{n}\mu
^{x-n-\gamma }\frac{\Gamma (y+n-\rho +1)}{(1+\tau )^{y+n-\rho +1}}$%
\end{tabular}%
\end{equation}%
We also need the posterior distribution function $F(\mu |x,y)$:%
\begin{equation}
\begin{tabular}{l}
$F(\mu |x,y)=\int_{0}^{\mu }f(t|x,y)dt=$ \\ 
$\sum_{n=0}^{x}\frac{\tau ^{y}}{m(x,y)x!y!}\binom{x}{n}\frac{\Gamma
(y+n-\rho +1)}{(1+\tau )^{y+n-\rho +1}}\int_{0}^{\mu }t^{x-n-\gamma
}e^{-t}dt=$ \\ 
$\sum_{n=0}^{x}\frac{\tau ^{y}}{m(x,y)x!y!}\binom{x}{n}\frac{\Gamma
(y+n-\rho +1)}{(1+\tau )^{y+n-\rho +1}}\Gamma (x-n-\gamma +1)\int_{0}^{\mu }%
\frac{1}{\Gamma (x-n-\gamma +1)}t^{(x-n-\gamma +1)-1}e^{-t}dt=$ \\ 
$\sum_{n=0}^{x}\frac{\tau ^{y}}{m(x,y)x!y!}\binom{x}{n}\frac{\Gamma
(y+n-\rho +1)}{(1+\tau )^{y+n-\rho +1}}\Gamma (x-n-\gamma +1)F_{\Gamma }(\mu
;x-n-\gamma +1,1)$%
\end{tabular}%
\end{equation}

where $F_{\Gamma }(.;\alpha ,\beta )$ is the distribution function of a
Gamma random variable with parameters $(\alpha ,\beta )$.

Now we need to extract an interval from the posterior distribution. We
will use the method of highest posterior density, which is the solution of
the system of equations%
\begin{equation}
\begin{tabular}{l}
$f(L|x,y)=f(U|x,y)$ \\ 
$F(U|x,y)-F(L|x,y)=1-\alpha $%
\end{tabular}%
\end{equation}

The advantage of this solution over the more common equal tail area solution
is that this method yields a smooth transition from one sided to two sided
intervals and therefore avoids the problem of flip-flopping, which is
further discussed in the section on Performance.

In the following we will denote the Bayesian methods by their priors, so for
example $\pi (\mu ,b)=1$ is the method with flat priors on both $\mu $ and $b
$, $\pi (\mu ,b)=1/\sqrt{\mu }$ uses Jeffrey's prior on $\mu $ and a flat
prior on $b$ etc.

\section{Performance of these Methods}

\subsection{\protect\bigskip Coverage}

The coverage\ $Cov$ of a set of confidence intervals $[L(x,y),U(x,y)]$ is
defined by 
\begin{equation}
\begin{tabular}{l}
$Cov(\mu ,b)=P\left( L(X,Y)\leq \mu \leq U(X,Y)|\mu ,b\right) =$ \\ 
$\sum_{x=0}^{\infty }\sum_{y=0}^{\infty }I_{[L(x,y),U(x,y)]}(\mu )f(x,y;\mu
,b)$%
\end{tabular}%
\end{equation}%
where $I_{A}(x)$ is the indicator function of the set \thinspace $A$ and $%
f(x,y;\mu ,b)$ is the probability density from equation \ref{pdf}.

A $(1-\alpha )100\%$ confidence interval is said to have coverage if for all 
$(\mu ,b)\in \lbrack 0,\infty )\times \lbrack 0,\infty )$ 
\begin{equation}
Cov(\mu ,b)\geq 1-\alpha
\end{equation}

For this paper we will restrict ourselves to the region of parameter space with 
$(\mu ,b)\in \lbrack 0,20]\times \lbrack 0,10]$, relevant for counting
experiments with low statistics. In the case of larger values of $\mu $ and $%
b$ one would likely use some asymptotic methods as studied by Cowan et.al. 
\cite{Cowan:2010js}.

The experimenter might on occasion decide ahead of time that they will only
calculate an upper limit, for example if theory suggests that the signal
rate is very small or even $0$. The distinction between upper limits and
two-sided confidence intervals is somewhat artificial because an upper limit
can always be viewed as a confidence interval with $L(x,y)=-\infty $ for all 
$x,y$. Moreover the property of coverage applies equally to both. One
important point, though, is that for a method that yields either one or the
other the experimenter must decide before seeing the data which one he wants
to use. Deciding this based on the observed data leads to the flip-flopping
problem, which generally leads to under-coverage. All of the methods used in
this paper have a smooth transition from upper limit to two-sided interval,
except for $CLs$, which by design yields upper limits only.

If at some point in parameter space we have $Cov(\mu ,b)>1-\alpha $ the
method is said to overcover. Overcoverage is \textquotedblleft legal\textquotedblright\ in the sense that a
method is still said to have coverage but is undesirable because it
generally comes at the price of larger intervals. Unfortunately in the case
of a discrete distribution such as the Poisson overcoverage at almost all
points in parameter space is unavoidable.

On the other hand if we have $Cov(\mu ,b)<1-\alpha $ the method is said to
undercover. This is a much bigger problem to the point of making the method
useless. \ In practice, though, a small amount of undercoverage is generally
deemed to be acceptable, and in fact as we shall see all the methods
described here undercover at least a little in some part of parameter space.
If one were to decide that any amount of undercoverage is unacceptable, one
could proceed as follows. Say the limits $[L(x,y),U(x,y)]$ have been
calculated to yield $90\%$ confidence intervals, but at some point $(\mu ,b)$
the actual coverage is only $85\%$. Then calculating the limits at a higher
nominal confidence level will also increase the actual worst coverage, and
there exists a nominal confidence level so that the actual lowest coverage
is the desired one.

This is known in Statistics as the method of adjusted p-values. For an
example see Aldor-Noiman et al. \cite{Aldor-Noiman} and for a general
discussion see Rolke and Buja \cite{Buja_calibrationfor}.

\begin{figure}[H]
\caption{Coverage and Expected Lengths for  
                  for the case $b=0.5$, $\protect\tau =1$ and $90\%$ confidence intervals}
\label{fig:05-1-90}
\centering
\includegraphics[width=0.98%
\textwidth,natwidth=450,natheight=450]{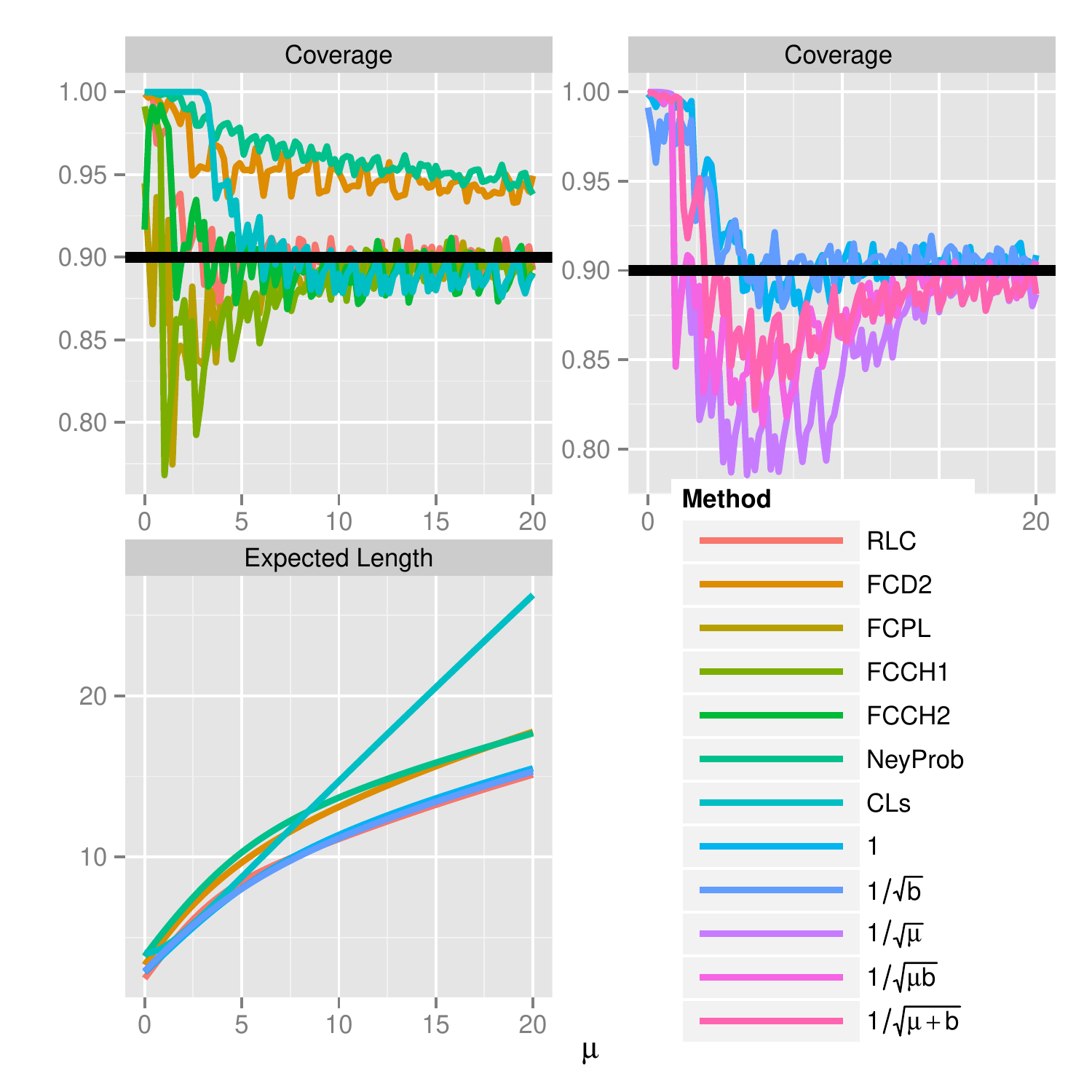}
\end{figure}

Let's begin with a graph of the coverage for the case $b=0.5$, $\tau =1$ and 
$90\%$ confidence intervals, shown in figure \ref{fig:05-1-90}. We see surprisingly bad performances of $FCCH1$
and $FCPL$. In the case of $FCCH1$ the minimum occurs for $\mu =1.0$ where
the actual coverage is only $77\%$. This turns out to be due to the fact
that for the case $x=0$, $y=0$ the upper limit of the 95\% interval of $FCCH1
$ is only $1.45$, so if we check coverage for (say) $\mu +b=1.5$ this case
is excluded, although $f(0,0;1.0,0.5)=0.12$, clearly the probability missing
for good coverage.

Similarly the (to) small limit of $FCPL$ for $x=0,y=0$ leads to bad
undercoverage, this time at $\mu =1.4$. The coverage of the Bayesian methods
with Jeffrey's prior on $\mu $ is also quite bad, mainly because these
priors favor smaller values of $\mu $. Finally the prior $\pi (\mu ,b)=1/%
\sqrt{\mu +b}$ leads to a method with coverage that is somewhat borderline. 

\begin{figure}[H]
\caption{Coverage and Expected Lengths for  
                  for the case $b=5$, $\protect\tau =1$ and $90\%$ confidence intervals}
\label{fig:5-1-90}
\centering
\includegraphics[width=0.98%
\textwidth,natwidth=450,natheight=450]{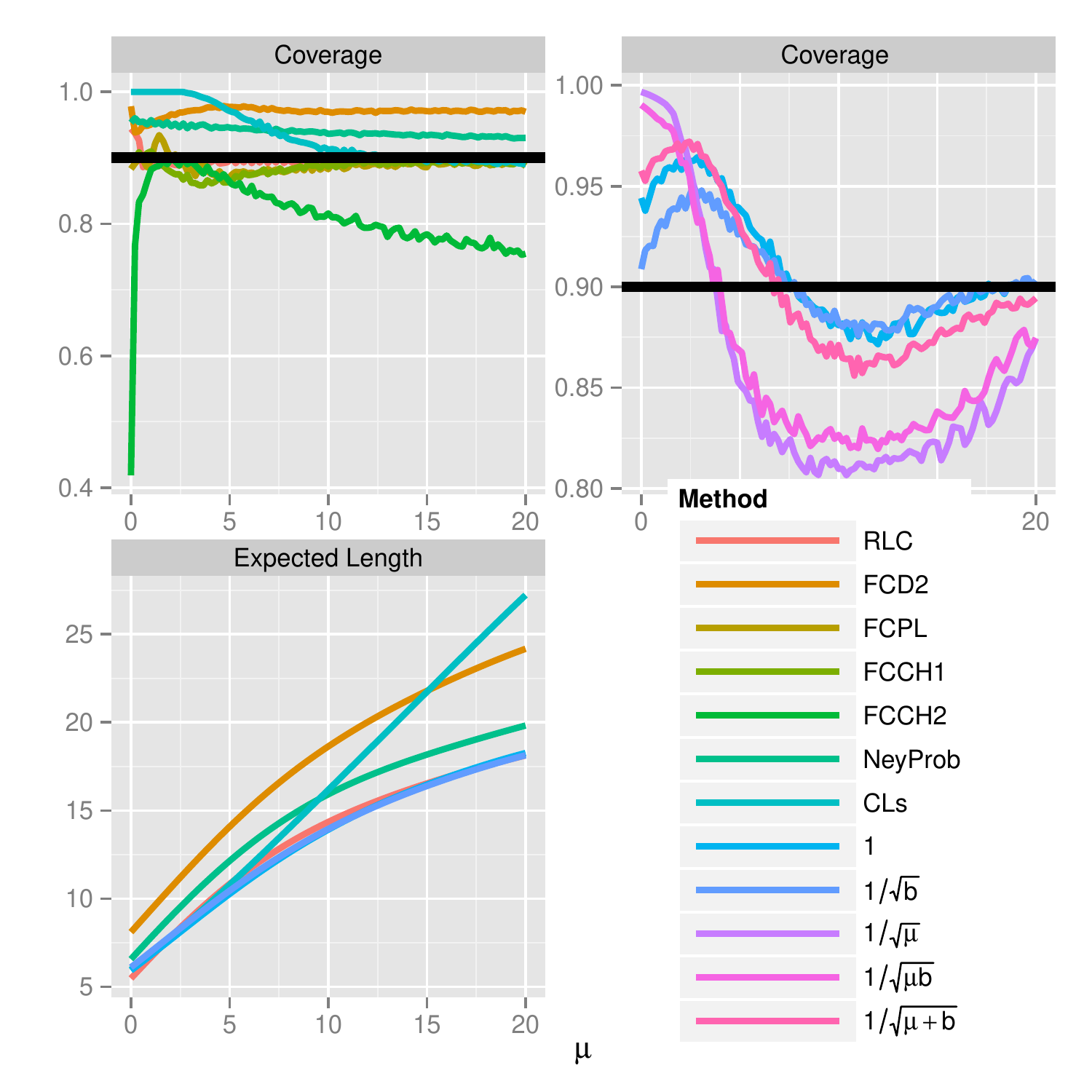}
\end{figure}

Figure \ref{fig:5-1-90} shows the coverage for
the case $b=5.0$, $\tau =1$ and $90\%$ confidence intervals. Here the worst
method is $FCCH2$, with a true coverage of $42\%$ if $\mu =0.0$! This is due
to the fact that the averaging over the lower limits leads very quickly to a
positive lower limit, for example if $x=2,y=0,\tau =1$ we get the $90\%$
interval $(0.09,4.29)$, and so for $\mu <0.9$ this case is rejected.

Finally we will find the worst coverage of each method by searching over a
grid on $0.5\leq b\leq 10$ and $0\leq \mu \leq 20$. The results are shown in
table \ref{tab:worst1.90}. As we saw before $FCCH1$, $FCCH2$, $FCPL$, $1/%
\sqrt{\mu}$, $1/\sqrt{\mu b}$ and $1/%
\sqrt{\mu +b}$ show some considerable undercoverage. These methods will therefore
be removed from further consideration.

Of course one should repeat the above studies for other values of $\tau $
and other confidence levels. In the appendix we have the
corresponding graphs and tables for the cases $\tau = 0.5, 2$ and $68\%$ and $95\%$
intervals. The results are very similar to those shown here.

\subsection{Other Considerations}

This leaves us with a choice of six methods. How do we decide among those?
Here the experimenter is free to use any criterion he wishes, provided that
the choice is made without consideration of the data. We already mentioned
two, namely avoiding the problem of flip-flopping and/or avoiding empty
intervals. All (except $CLs$, which always yields upper limits) of the
methods discussed here have a smooth transition from upper limits to
two-sided intervals, and so flip-flopping is not an issue. The only method
which could yield empty intervals is $NeyProb$, which might be eliminated
from consideration for that reason.

A very common criterion for the performance of confidence intervals in
Statistics is the expected mean length, defined by%
\begin{equation}
EL(\mu ,b)=\sum_{x=0}^{\infty }\sum_{y=0}^{\infty }\left[ U(x,y)-L(x,y)%
\right] f(x,y;\mu ,b)
\end{equation}

Why short intervals are desirable is most easily seen in the case of upper
limits, where it simply means a tighter bound on the parameter of interest. Even in the case of a
two-sided interval, knowing that the parameter is most likely in the
interval (say) $(2.5,4.9)$ is better than just knowing it is in the interval 
$(2.1,5.3)$. In general, for confidence intervals the expected length plays a
role similar to the concept of power in the case of hypothesis testing. 

It should be noted that the expected length is metric dependent. So a change in the parametrization of the problem might also change which method yields shortest expected length.

In figure \ref{fig:05-1-90} we show the expected length as a
function of $\mu $ for the case $b=0.5$, $\tau =1$ and $90\%$ confidence
intervals, and in figure \ref{fig:5-1-90} the same for the case $b=5.0$%
. In both cases $RLC$ has the shortest intervals for small $\mu $ and the Bayesian methods for larger ones. As one would expect the overcoverage observed
for $CLs$, $NeyProb$ and $FC2D$ leads to larger expected intervals
\section{Conclusions}

We have studied the coverage and the expected length of the confidence
intervals generated by a number of methods for the \textquotedblleft On-Off \textquotedblright\ problem. The
intervals were derived using a variety of methodologies and include all
those in common use today. We find that the $RLC$ limits based on the
profile likelihood and the limits derived using the Bayesian methodology
with a flat prior on the signal rate $\mu $ are best, all having acceptable coverage and shortest
expected length. It is noteworthy that the oldest method in this study,
namely the method implemented in MINUIT/MINOS, is still a strong contender even today,
at least when used with some adjustments for the cases when $x<y/\tau $ as is
done in $RLC$. It should also be mentioned that just because a flat prior on $\mu $
leads to methods with good coverage for the \textquotedblleft On-Off \textquotedblright\ problem studied here, this does not necessarily mean that flat priors will always be the best
choice. 

This study was possible because for this simple model we were able to find
explicit formulas for the limits. It would obviously be desirable to do
similar studies for more complicated models, for example including
uncertainties in $\tau $, including efficiencies with their uncertainties as
well extensions such as multiple channels. In those cases, though, deriving
explicit formulas will be difficult if not impossible, and as soon as MC
methods are needed to calculate limits the scope of any coverage study will
be severely limited. Nevertheless we hope this study will provide some
guidance as to which methods are most promising.

\bibliographystyle{plain}
\bibliography{rolke}
{}

\newpage

\section{Appendix}
\subsection{Confidence region for Feldman-Cousins}

\subsection{Graphs and Tables of Coverage and Expected Lengths}

\subsubsection{Case $\protect\tau =0.5 $ and $68\%$ Confidence Intervals}
\begin{table}[tbh]
\caption{Worst coverage of each method, for $0.5\leq b\leq 10$ and $0\leq 
        \protect\mu \leq 20$. $\protect\tau =$ 0.5 $ and $ 68 $\%$ confidence intervals}
\label{tab:worst0.5.68}
\begin{tabular}{|c|c|c|c|c|c|c|c|}
\hline
&    RLC   &     FCD2   &   FCPL &   FCCH1   &   FCCH2   &    NeyProb   &     CLs  \\
\hline
\hline
b  & $ 3 $ & $ 0.5 $ & $ 0.5 $ & $ 0.5 $ & $ 10 $ & $ 10 $ & $ 0.5 $  \\
\hline
$\protect\mu $  & $ 8.6 $ & $ 8.2 $ & $ 1.2 $ & $ 0.8 $ & $ 20 $ & $ 19.6 $ & $ 6.5 $  \\
\hline
Coverage    & $ 63.7 $ & $ 86.2 $ & $ 81.3 $ & $ 72.2 $ & $ 24.3 $ & $ 81.6 $ & $ 51.8 $  \\
\hline
\end{tabular}
\begin{tabular}{|c|c|c|c|c||c|}
\hline
&  $1$ &  $1/\sqrt{\mu }$ & $1/\sqrt{b}$ & $1/\sqrt{\mu b }$ & $1/\sqrt{\mu +b}$ \\
\hline
\hline
b  & $ 0.5 $ & $ 10 $ & $ 0.5 $ & $ 0.5 $ & $ 0.5 $  \\
\hline
$\protect\mu $  & $ 4.1 $ & $ 9 $ & $ 4.5 $ & $ 0 $ & $ 3.3 $  \\
\hline
Coverage    & $ 59 $ & $ 61.9 $ & $ 33.5 $ & $ 39.3 $ & $ 50.2 $  \\
\hline
\end{tabular}
\end{table}

\begin{figure}[H]
\caption{Coverage and Expected Lengths for  
                  for the case $b=0.5$, $\protect\tau =0.5$ and $68\%$ confidence intervals}
\label{fig:05-05-68}
\centering
\includegraphics[width=0.98%
\textwidth,natwidth=450,natheight=450]{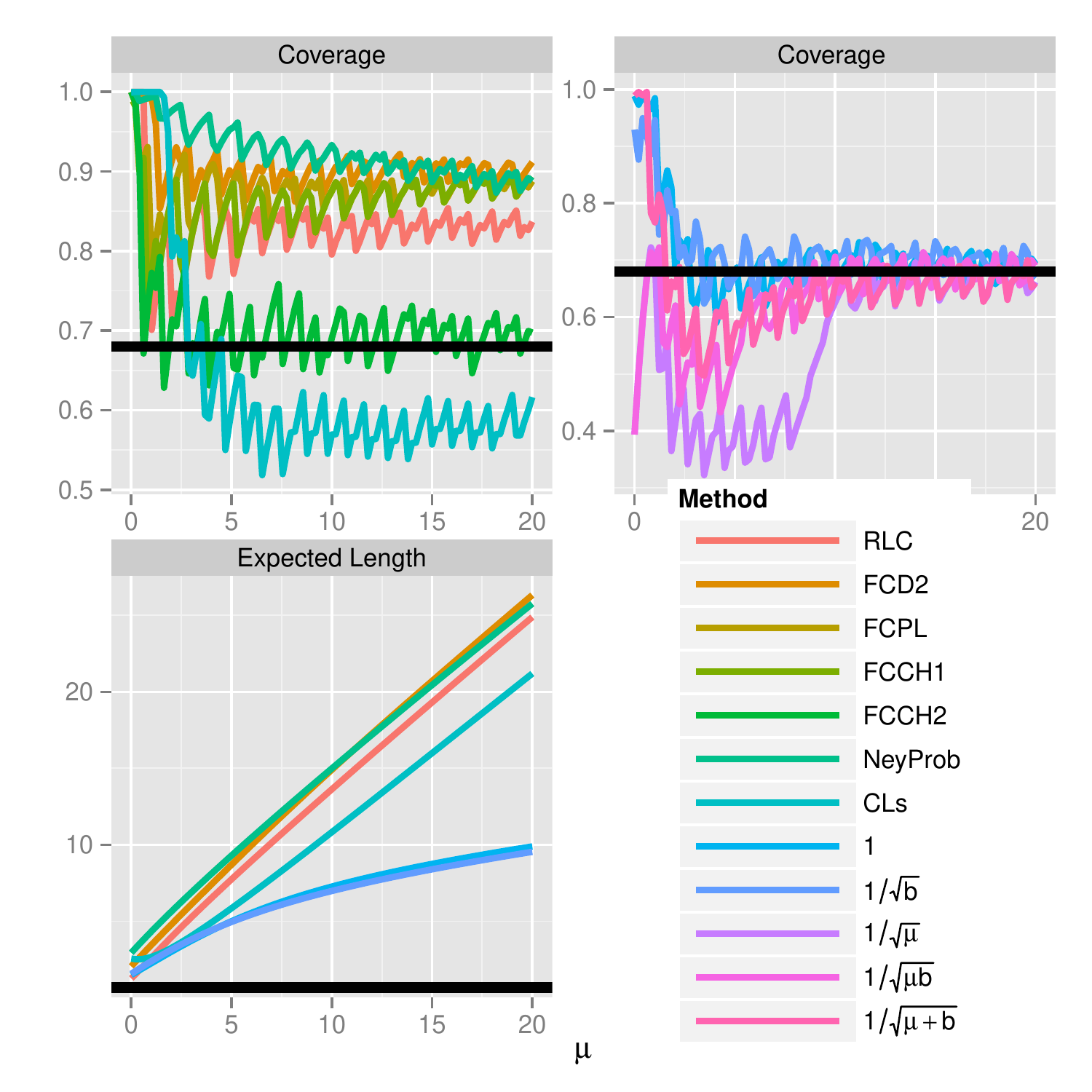}
\end{figure}
\begin{figure}[H]
\caption{Coverage and Expected Lengths for  
                  for the case $b=3$, $\protect\tau =0.5$ and $68\%$ confidence intervals}
\label{fig:3-05-68}
\centering
\includegraphics[width=0.98%
\textwidth,natwidth=450,natheight=450]{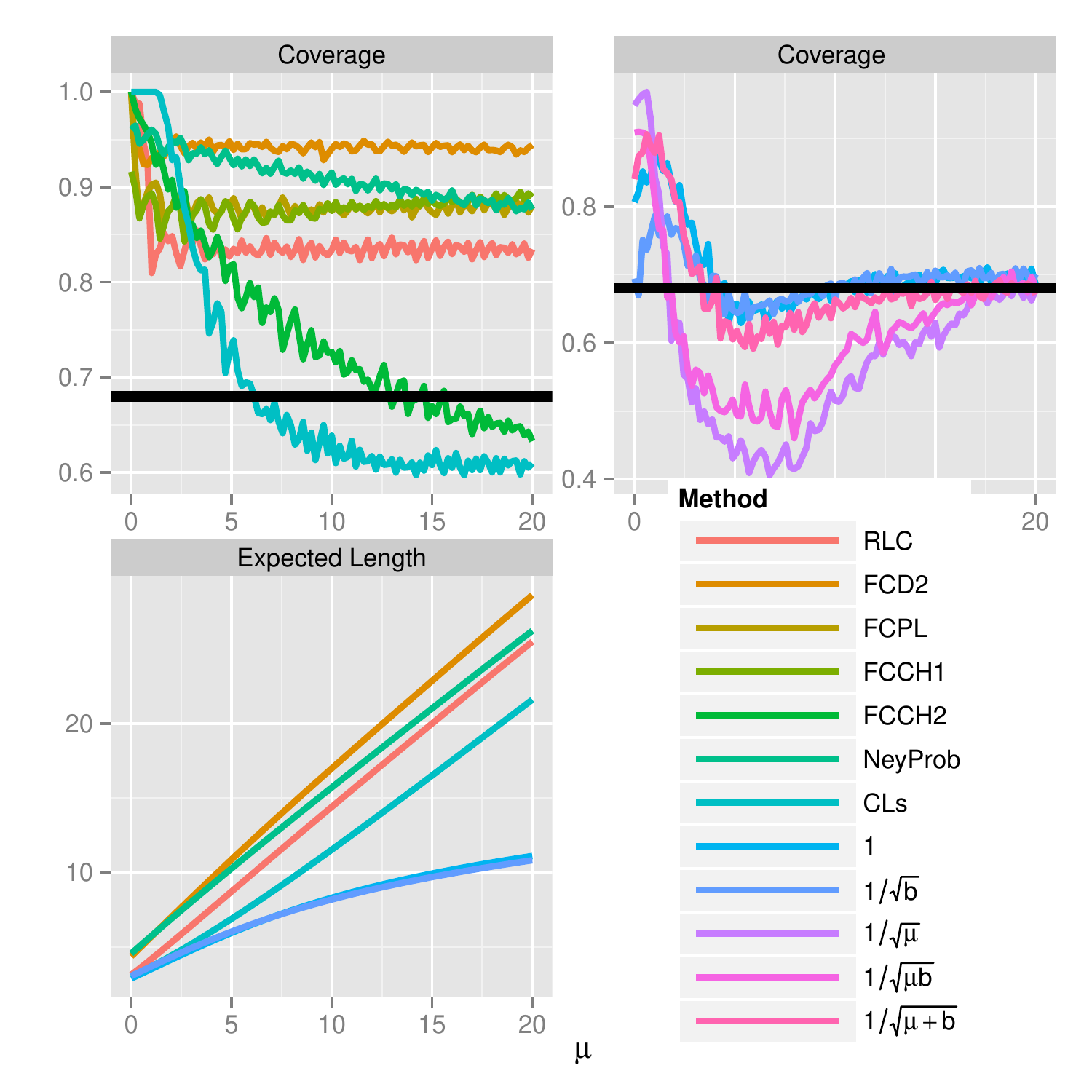}
\end{figure}
\begin{figure}[H]
\caption{Coverage and Expected Lengths for  
                  for the case $b=5$, $\protect\tau =0.5$ and $68\%$ confidence intervals}
\label{fig:5-05-68}
\centering
\includegraphics[width=0.98%
\textwidth,natwidth=450,natheight=450]{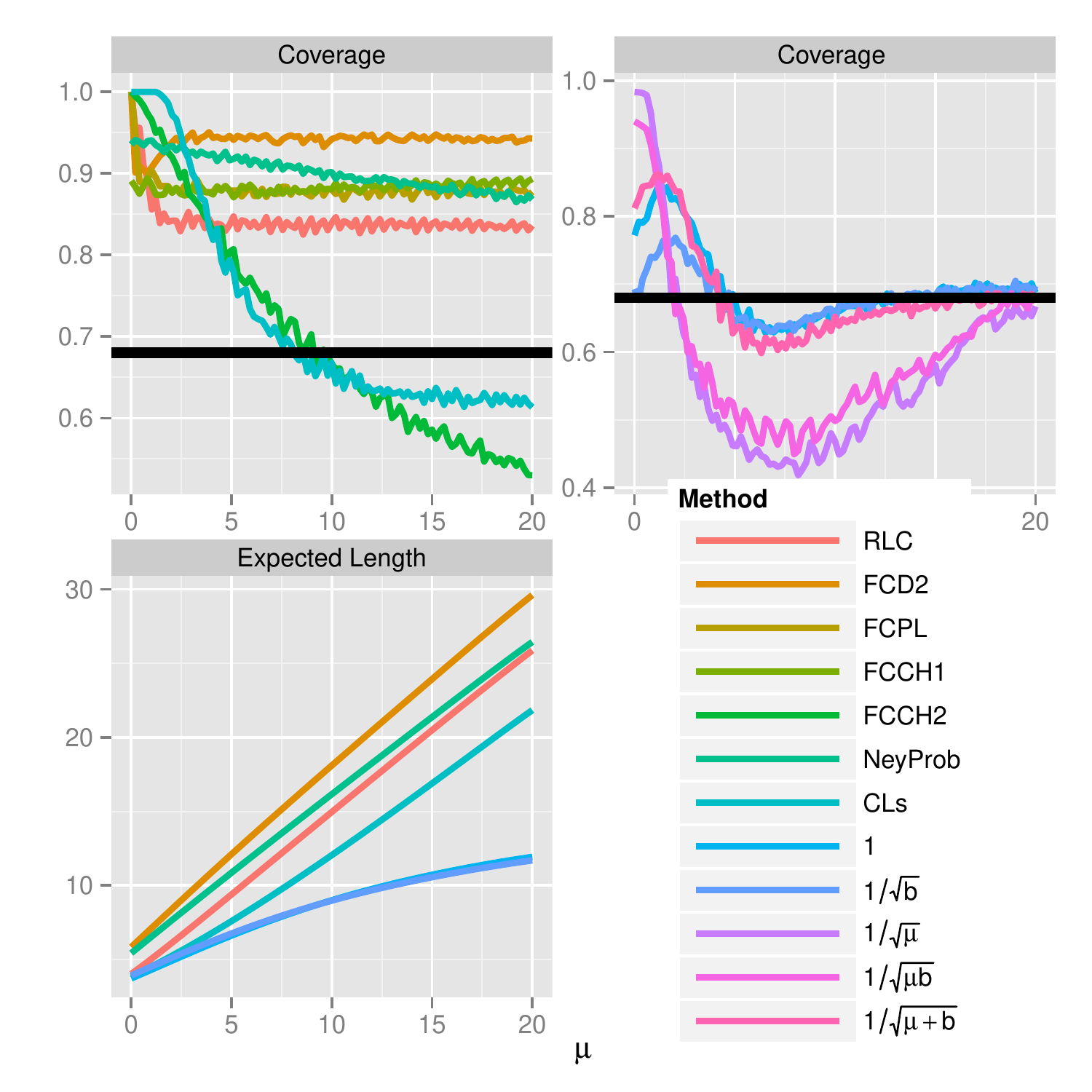}
\end{figure}

\subsubsection{Case $\protect\tau =0.5 $ and $90\%$ Confidence Intervals}
\begin{table}[tbh]
\caption{Worst coverage of each method, for $0.5\leq b\leq 10$ and $0\leq 
        \protect\mu \leq 20$. $\protect\tau =$ 0.5 $ and $ 90 $\%$ confidence intervals}
\label{tab:worst0.5.90}
\begin{tabular}{|c|c|c|c|c|c|c|c|}
\hline
&    RLC   &     FCD2   &   FCPL &   FCCH1   &   FCCH2   &    NeyProb   &     CLs  \\
\hline
\hline
b  & $ 10 $ & $ 10 $ & $ 3.2 $ & $ 0.5 $ & $ 10 $ & $ 10 $ & $ 0.5 $  \\
\hline
$\protect\mu $  & $ 20 $ & $ 0.4 $ & $ 0.4 $ & $ 1.6 $ & $ 0 $ & $ 20 $ & $ 16.3 $  \\
\hline
Coverage    & $ 87.8 $ & $ 91.8 $ & $ 87.8 $ & $ 87.4 $ & $ 21.1 $ & $ 90.3 $ & $ 83.8 $  \\
\hline
\end{tabular}
\begin{tabular}{|c|c|c|c|c||c|}
\hline
&  $1$ &  $1/\sqrt{\mu }$ & $1/\sqrt{b}$ & $1/\sqrt{\mu b }$ & $1/\sqrt{\mu +b}$ \\
\hline
\hline
b  & $ 0.5 $ & $ 10 $ & $ 0.5 $ & $ 0.5 $ & $ 0.5 $  \\
\hline
$\protect\mu $  & $ 8.2 $ & $ 18.8 $ & $ 9.4 $ & $ 6.5 $ & $ 7.8 $  \\
\hline
Coverage    & $ 83.3 $ & $ 86.1 $ & $ 73.8 $ & $ 78.7 $ & $ 79.8 $  \\
\hline
\end{tabular}
\end{table}

\begin{figure}[H]
\caption{Coverage and Expected Lengths for  
                  for the case $b=0.5$, $\protect\tau =0.5$ and $90\%$ confidence intervals}
\label{fig:05-05-90}
\centering
\includegraphics[width=0.98%
\textwidth,natwidth=450,natheight=450]{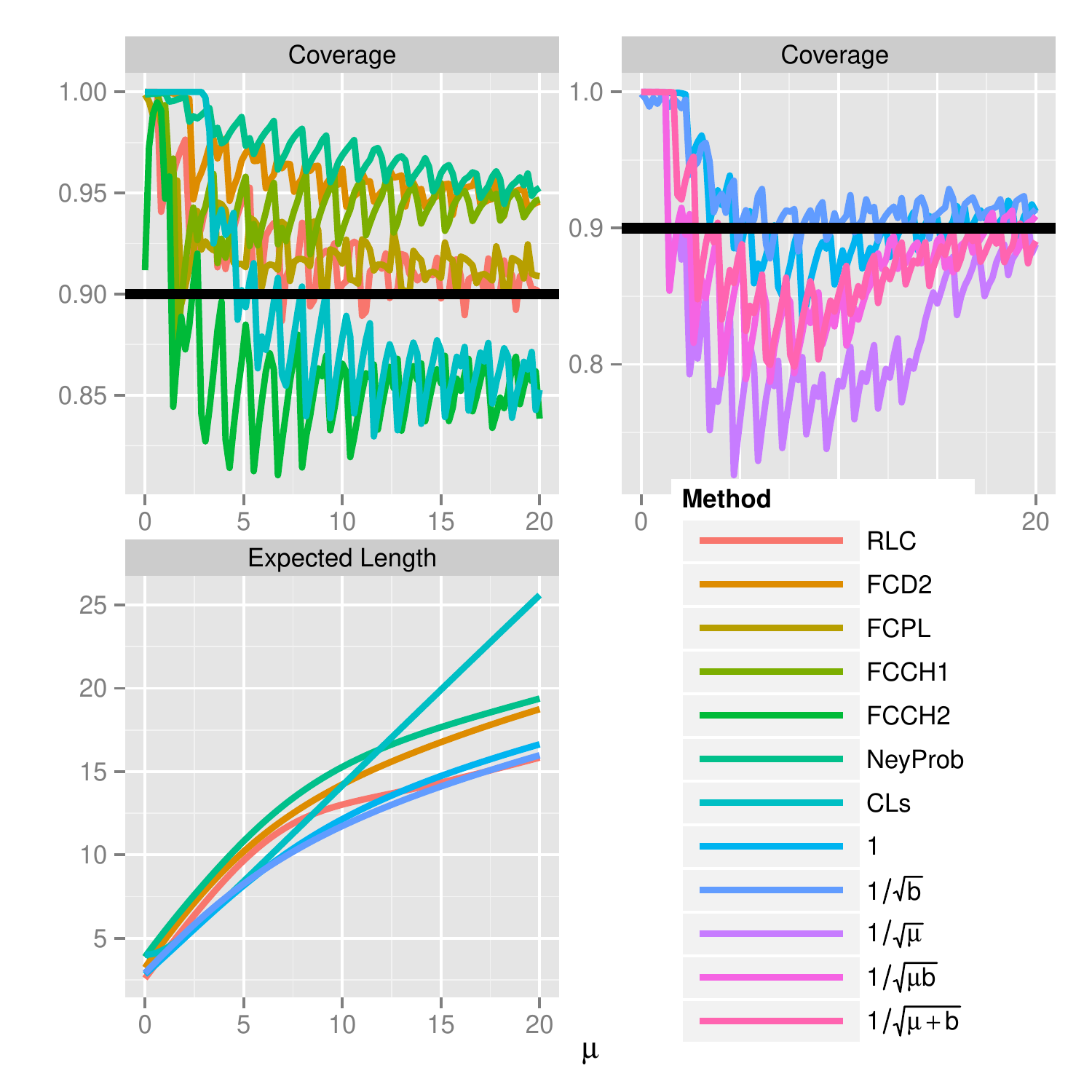}
\end{figure}
\begin{figure}[H]
\caption{Coverage and Expected Lengths for  
                  for the case $b=3$, $\protect\tau =0.5$ and $90\%$ confidence intervals}
\label{fig:3-05-90}
\centering
\includegraphics[width=0.98%
\textwidth,natwidth=450,natheight=450]{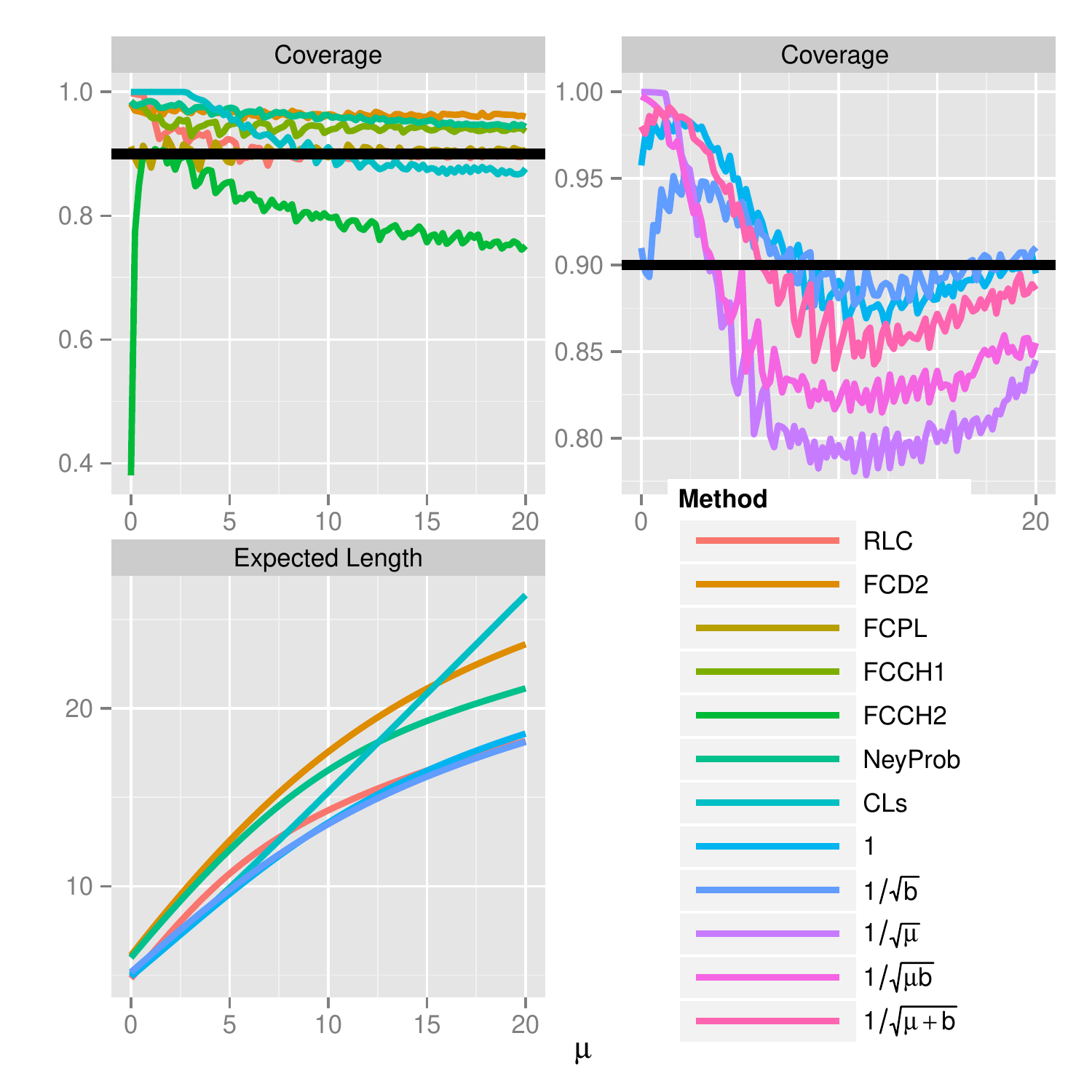}
\end{figure}
\begin{figure}[H]
\caption{Coverage and Expected Lengths for  
                  for the case $b=5$, $\protect\tau =0.5$ and $90\%$ confidence intervals}
\label{fig:5-05-90}
\centering
\includegraphics[width=0.98%
\textwidth,natwidth=450,natheight=450]{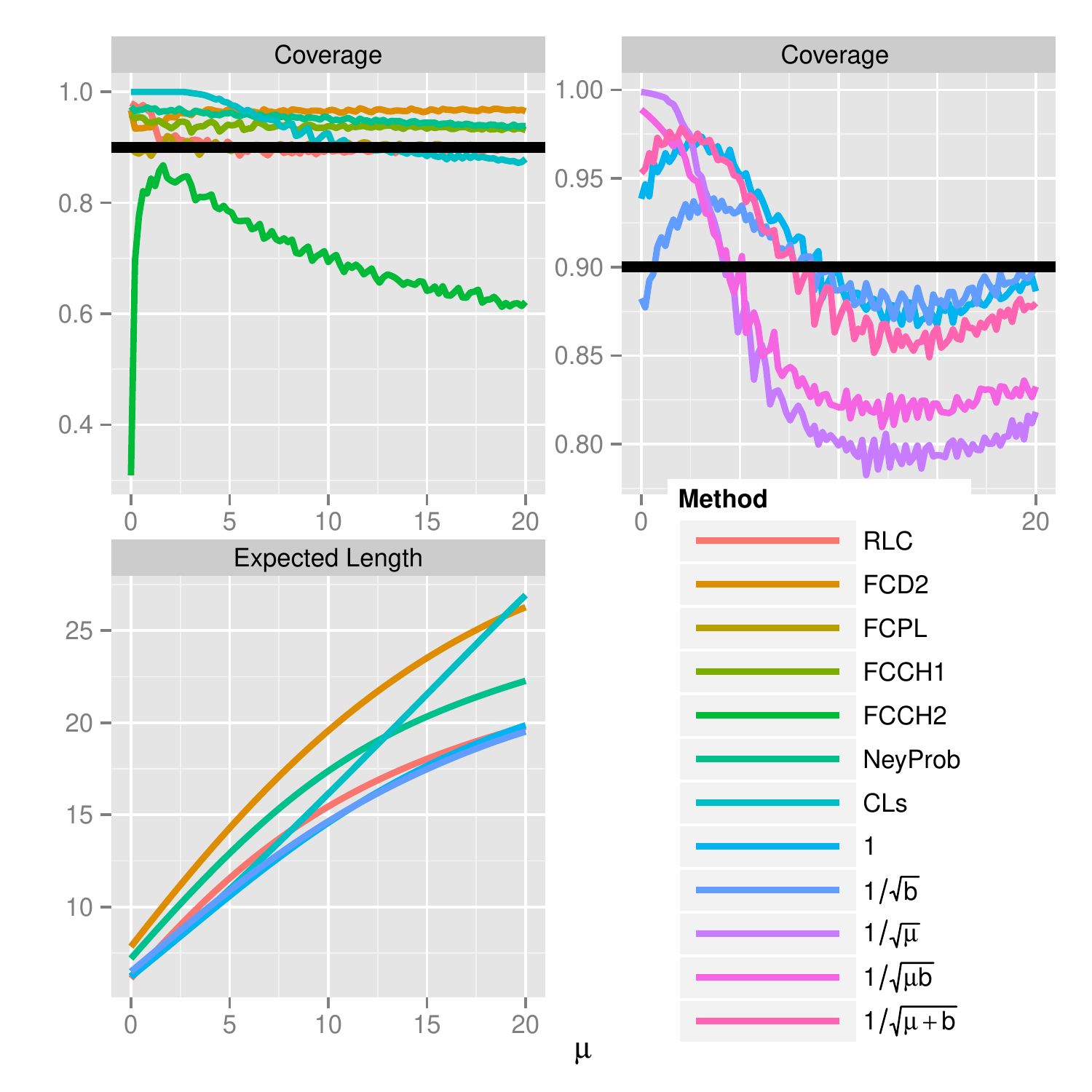}
\end{figure}

\subsubsection{Case $\protect\tau =0.5 $ and $95\%$ Confidence Intervals}
\begin{table}[tbh]
\caption{Worst coverage of each method, for $0.5\leq b\leq 10$ and $0\leq 
        \protect\mu \leq 20$. $\protect\tau =$ 0.5 $ and $ 95 $\%$ confidence intervals}
\label{tab:worst0.5.95}
\begin{tabular}{|c|c|c|c|c|c|c|c|}
\hline
&    RLC   &     FCD2   &   FCPL &   FCCH1   &   FCCH2   &    NeyProb   &     CLs  \\
\hline
\hline
b  & $ 10 $ & $ 9.8 $ & $ 10 $ & $ 10 $ & $ 10 $ & $ 10 $ & $ 0.5 $  \\
\hline
$\protect\mu $  & $ 20 $ & $ 0.4 $ & $ 20 $ & $ 20 $ & $ 0 $ & $ 20 $ & $ 11.8 $  \\
\hline
Coverage    & $ 92.6 $ & $ 94.8 $ & $ 92.6 $ & $ 93.6 $ & $ 25.6 $ & $ 93.6 $ & $ 90.9 $  \\
\hline
\end{tabular}
\begin{tabular}{|c|c|c|c|c||c|}
\hline
&  $1$ &  $1/\sqrt{\mu }$ & $1/\sqrt{b}$ & $1/\sqrt{\mu b }$ & $1/\sqrt{\mu +b}$ \\
\hline
\hline
b  & $ 10 $ & $ 10 $ & $ 0.5 $ & $ 10 $ & $ 0.5 $  \\
\hline
$\protect\mu $  & $ 20 $ & $ 20 $ & $ 6.9 $ & $ 19.6 $ & $ 11.4 $  \\
\hline
Coverage    & $ 90.9 $ & $ 91.6 $ & $ 83.8 $ & $ 88.2 $ & $ 88.8 $  \\
\hline
\end{tabular}
\end{table}

\begin{figure}[H]
\caption{Coverage and Expected Lengths for  
                  for the case $b=0.5$, $\protect\tau =0.5$ and $95\%$ confidence intervals}
\label{fig:05-05-95}
\centering
\includegraphics[width=0.98%
\textwidth,natwidth=450,natheight=450]{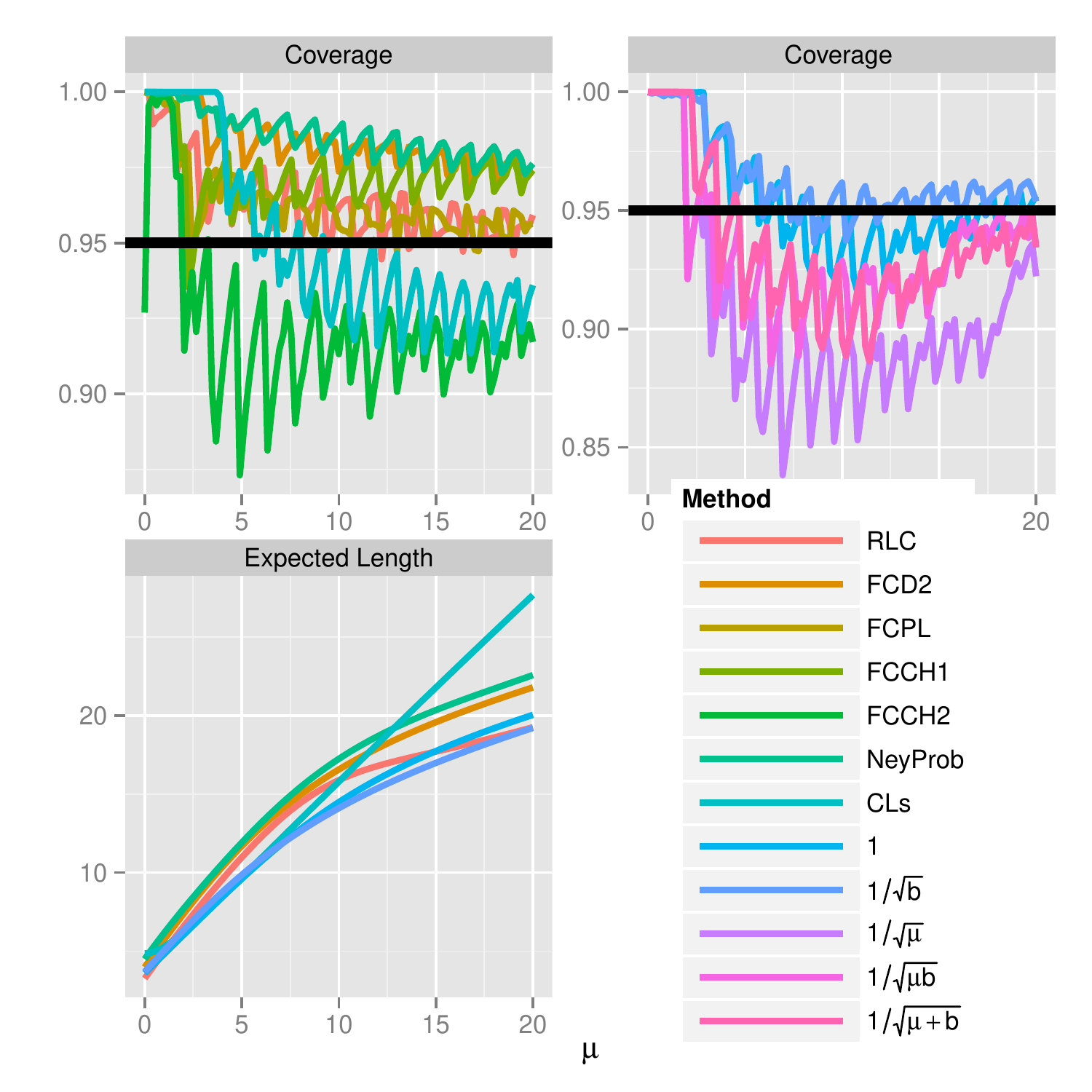}
\end{figure}
\begin{figure}[H]
\caption{Coverage and Expected Lengths for  
                  for the case $b=3$, $\protect\tau =0.5$ and $95\%$ confidence intervals}
\label{fig:3-05-95}
\centering
\includegraphics[width=0.98%
\textwidth,natwidth=450,natheight=450]{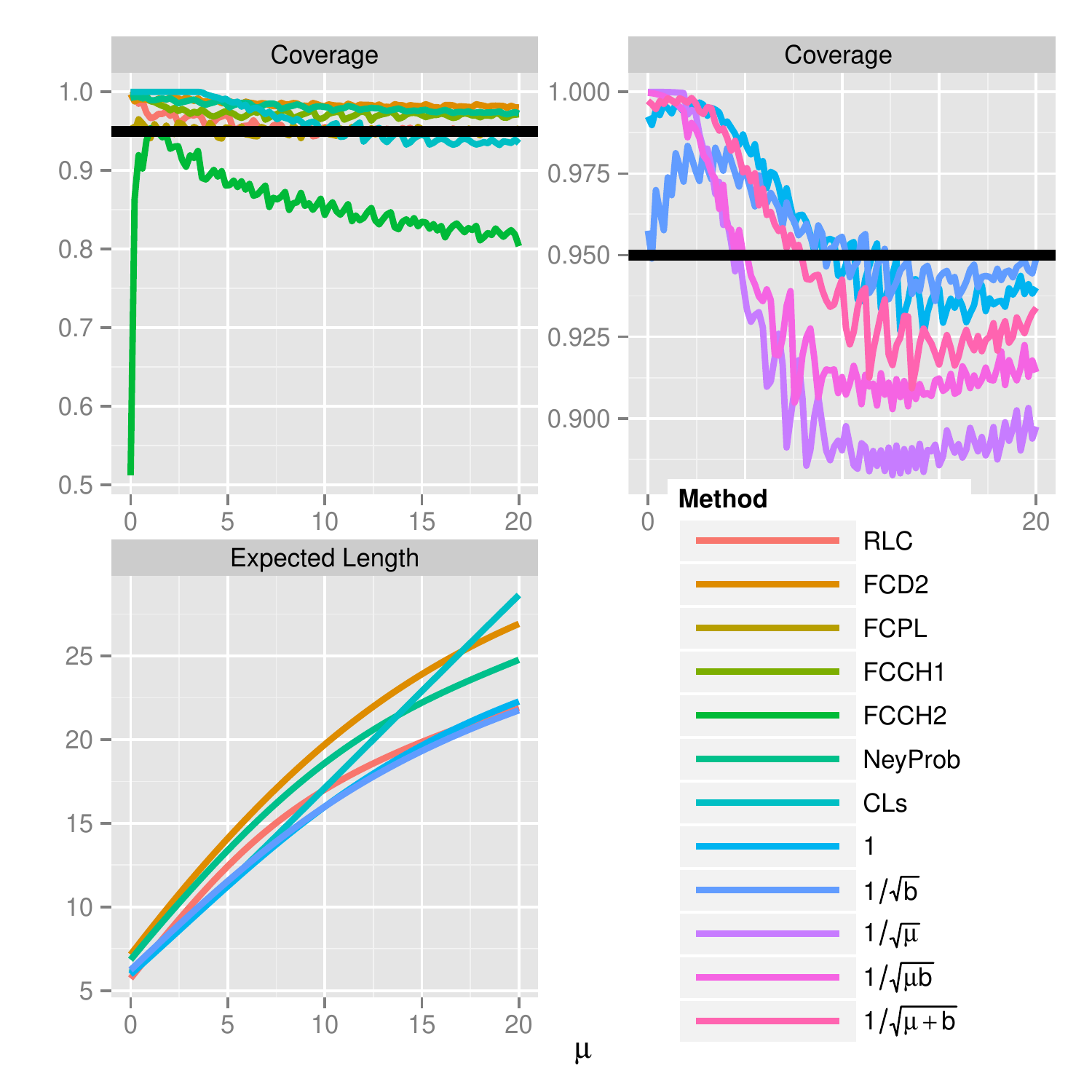}
\end{figure}
\begin{figure}[H]
\caption{Coverage and Expected Lengths for  
                  for the case $b=5$, $\protect\tau =0.5$ and $95\%$ confidence intervals}
\label{fig:5-05-95}
\centering
\includegraphics[width=0.98%
\textwidth,natwidth=450,natheight=450]{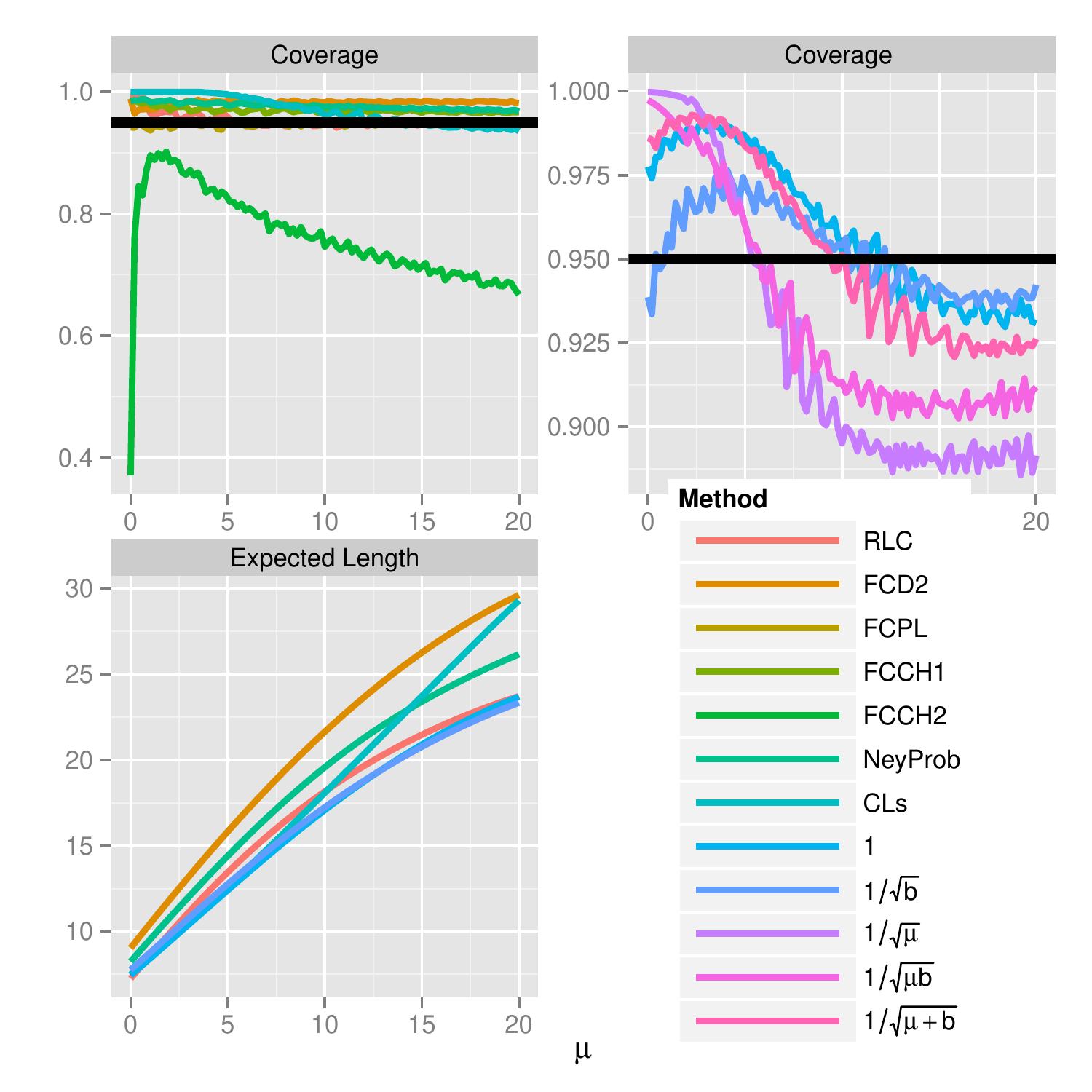}
\end{figure}

\subsubsection{Case $\protect\tau =1 $ and $68\%$ Confidence Intervals}
\begin{table}[tbh]
\caption{Worst coverage of each method, for $0.5\leq b\leq 10$ and $0\leq 
        \protect\mu \leq 20$. $\protect\tau =$ 1 $ and $ 68 $\%$ confidence intervals}
\label{tab:worst1.68}
\begin{tabular}{|c|c|c|c|c|c|c|c|}
\hline
&    RLC   &     FCD2   &   FCPL &   FCCH1   &   FCCH2   &    NeyProb   &     CLs  \\
\hline
\hline
b  & $ 1.3 $ & $ 0.5 $ & $ 0.5 $ & $ 0.5 $ & $ 10 $ & $ 10 $ & $ 0.5 $  \\
\hline
$\protect\mu $  & $ 1.6 $ & $ 4.5 $ & $ 0.8 $ & $ 0.4 $ & $ 0 $ & $ 20 $ & $ 7.3 $  \\
\hline
Coverage    & $ 63.6 $ & $ 79 $ & $ 48.3 $ & $ 43.1 $ & $ 18 $ & $ 72.8 $ & $ 60.9 $  \\
\hline
\end{tabular}
\begin{tabular}{|c|c|c|c|c||c|}
\hline
&  $1$ &  $1/\sqrt{\mu }$ & $1/\sqrt{b}$ & $1/\sqrt{\mu b }$ & $1/\sqrt{\mu +b}$ \\
\hline
\hline
b  & $ 10 $ & $ 8.6 $ & $ 0.5 $ & $ 0.5 $ & $ 0.5 $  \\
\hline
$\protect\mu $  & $ 7.8 $ & $ 6.9 $ & $ 0 $ & $ 2.9 $ & $ 2.4 $  \\
\hline
Coverage    & $ 60.9 $ & $ 61.8 $ & $ 39.3 $ & $ 37.5 $ & $ 51.1 $  \\
\hline
\end{tabular}
\end{table}

\begin{figure}[H]
\caption{Coverage and Expected Lengths for  
                  for the case $b=0.5$, $\protect\tau =1$ and $68\%$ confidence intervals}
\label{fig:05-1-68}
\centering
\includegraphics[width=0.98%
\textwidth,natwidth=450,natheight=450]{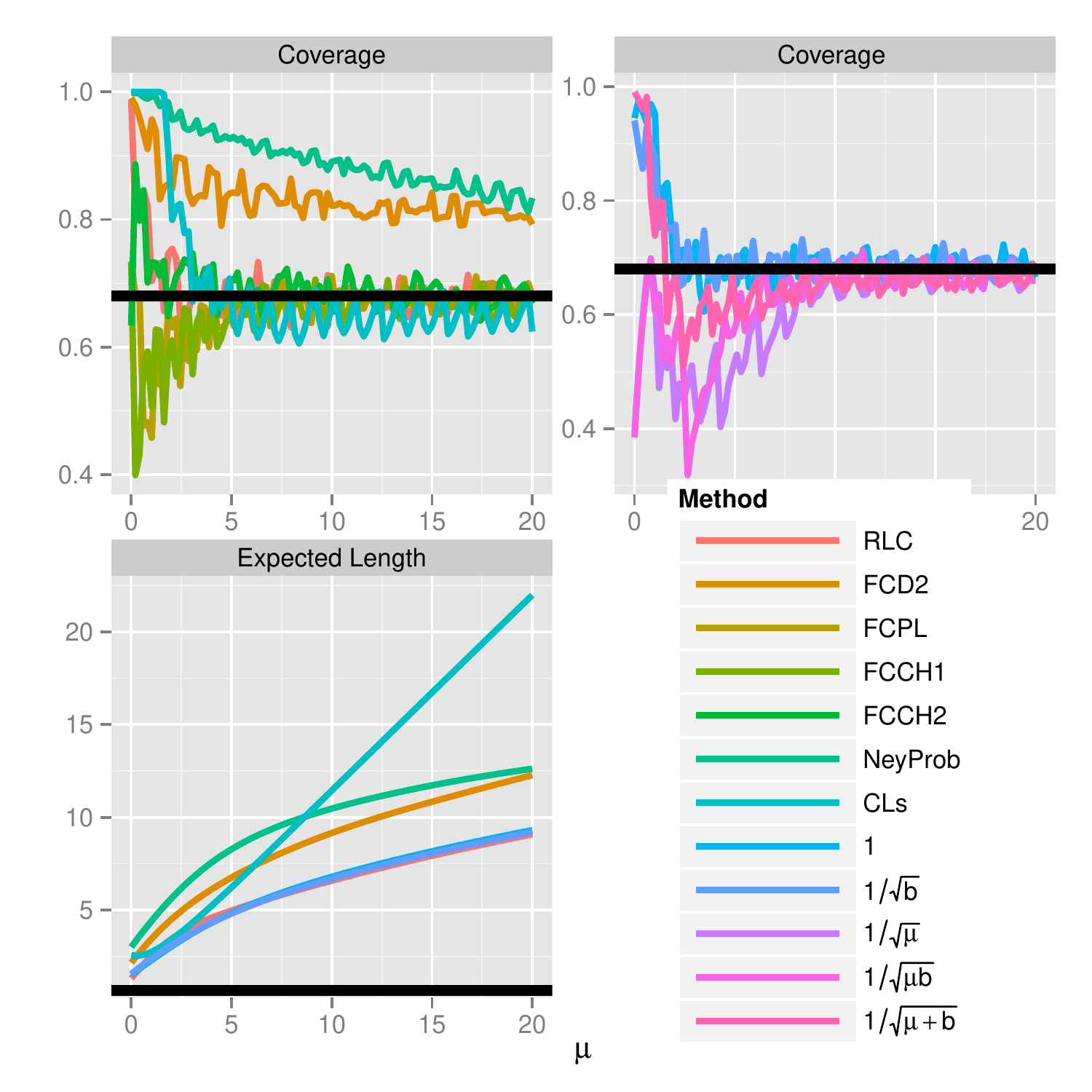}
\end{figure}
\begin{figure}[H]
\caption{Coverage and Expected Lengths for  
                  for the case $b=3$, $\protect\tau =1$ and $68\%$ confidence intervals}
\label{fig:3-1-68}
\centering
\includegraphics[width=0.98%
\textwidth,natwidth=450,natheight=450]{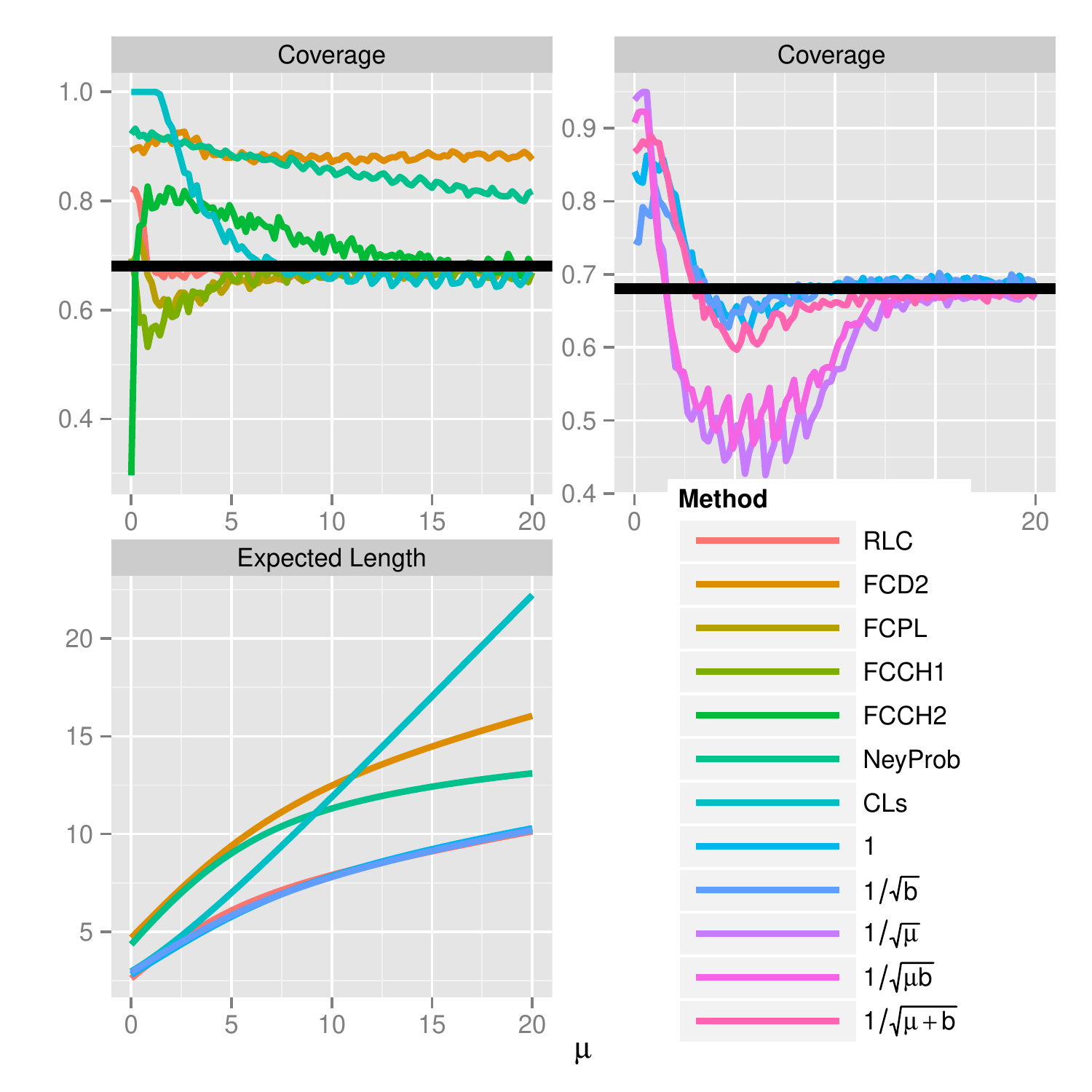}
\end{figure}
\begin{figure}[H]
\caption{Coverage and Expected Lengths for  
                  for the case $b=5$, $\protect\tau =1$ and $68\%$ confidence intervals}
\label{fig:5-1-68}
\centering
\includegraphics[width=0.98%
\textwidth,natwidth=450,natheight=450]{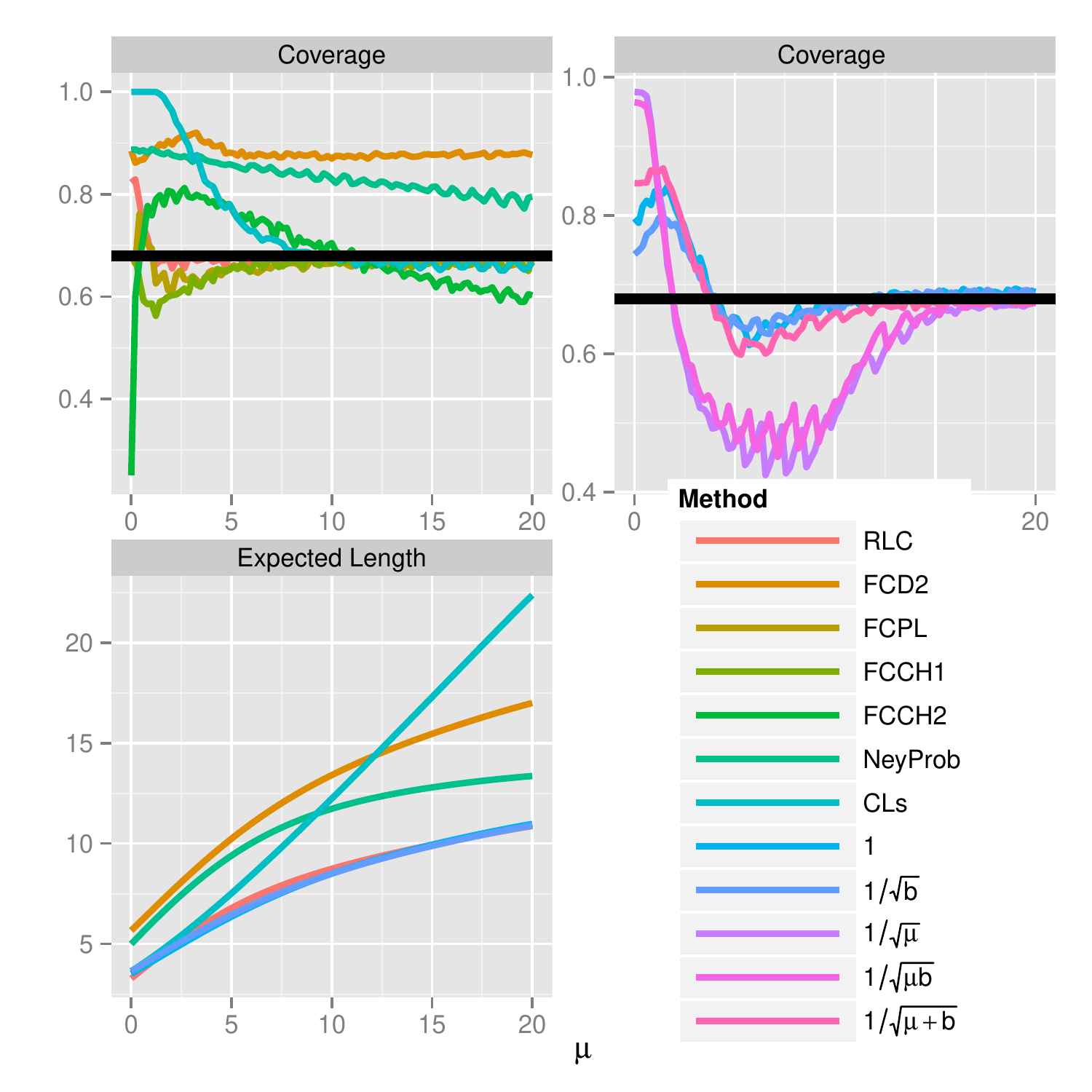}
\end{figure}

\subsubsection{Case $\protect\tau =1 $ and $90\%$ Confidence Intervals}
\begin{table}[tbh]
\caption{Worst coverage of each method, for $0.5\leq b\leq 10$ and $0\leq 
        \protect\mu \leq 20$. $\protect\tau =$ 1 $ and $ 90 $\%$ confidence intervals}
\label{tab:worst1.90}
\begin{tabular}{|c|c|c|c|c|c|c|c|}
\hline
&    RLC   &     FCD2   &   FCPL &   FCCH1   &   FCCH2   &    NeyProb   &     CLs  \\
\hline
\hline
b  & $ 0.5 $ & $ 10 $ & $ 0.7 $ & $ 0.5 $ & $ 10 $ & $ 10 $ & $ 10 $  \\
\hline
$\protect\mu $  & $ 3.7 $ & $ 0.4 $ & $ 2.9 $ & $ 1.2 $ & $ 20 $ & $ 20 $ & $ 20 $  \\
\hline
Coverage    & $ 87.7 $ & $ 88 $ & $ 83.4 $ & $ 80 $ & $ 28.8 $ & $ 90.5 $ & $ 86.8 $  \\
\hline
\end{tabular}
\begin{tabular}{|c|c|c|c|c||c|}
\hline
&  $1$ &  $1/\sqrt{\mu }$ & $1/\sqrt{b}$ & $1/\sqrt{\mu b }$ & $1/\sqrt{\mu +b}$ \\
\hline
\hline
b  & $ 2.1 $ & $ 10 $ & $ 0.5 $ & $ 10 $ & $ 0.5 $  \\
\hline
$\protect\mu $  & $ 9.8 $ & $ 14.7 $ & $ 6.9 $ & $ 20 $ & $ 6.9 $  \\
\hline
Coverage    & $ 86.9 $ & $ 87.4 $ & $ 80.4 $ & $ 81.3 $ & $ 83.9 $  \\
\hline
\end{tabular}
\end{table}

\begin{figure}[H]
\caption{Coverage and Expected Lengths for  
                  for the case $b=3$, $\protect\tau =1$ and $90\%$ confidence intervals}
\label{fig:3-1-90}
\centering
\includegraphics[width=0.98%
\textwidth,natwidth=450,natheight=450]{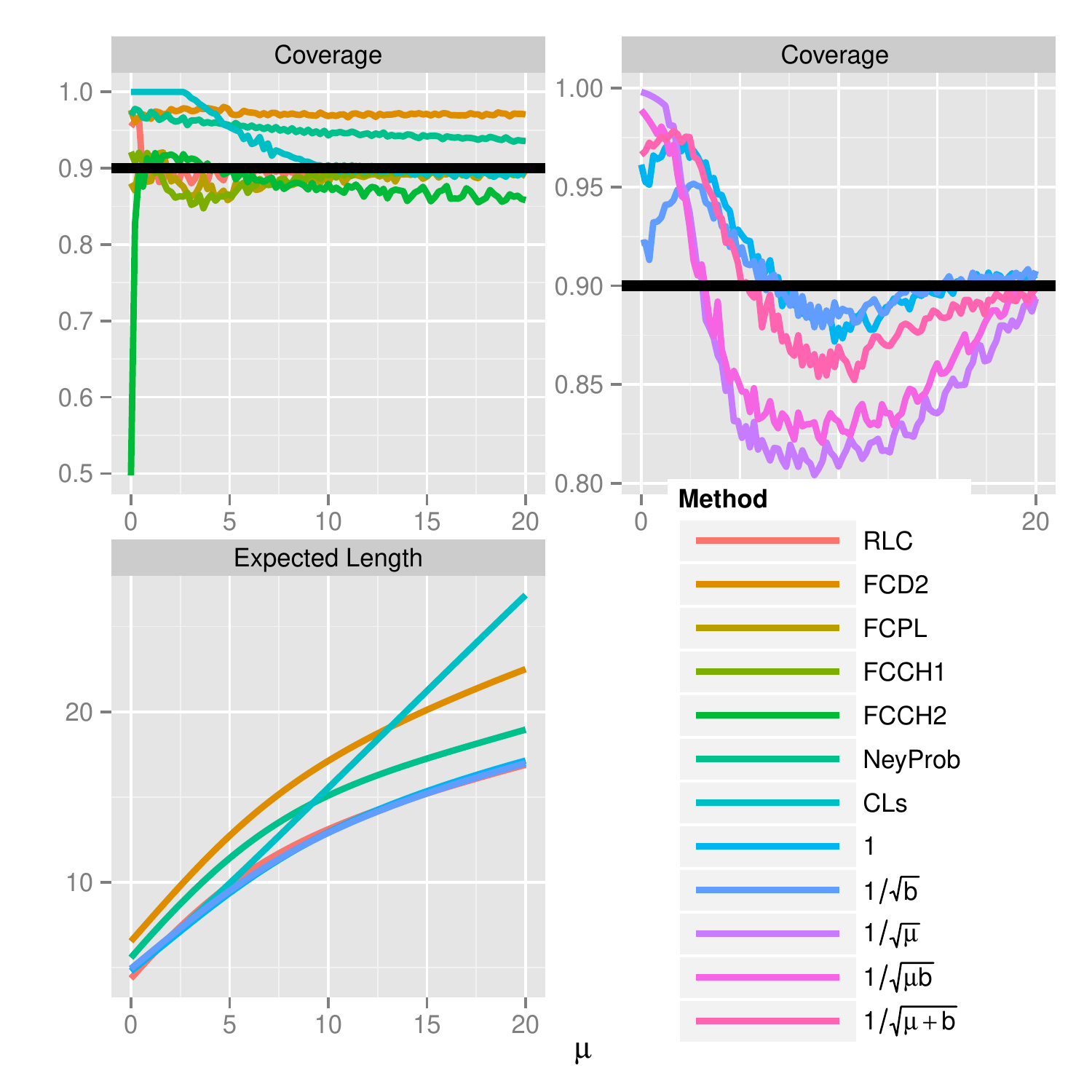}
\end{figure}

\subsubsection{Case $\protect\tau =1 $ and $95\%$ Confidence Intervals}
\begin{table}[tbh]
\caption{Worst coverage of each method, for $0.5\leq b\leq 10$ and $0\leq 
        \protect\mu \leq 20$. $\protect\tau =$ 1 $ and $ 95 $\%$ confidence intervals}
\label{tab:worst1.95}
\begin{tabular}{|c|c|c|c|c|c|c|c|}
\hline
&    RLC   &     FCD2   &   FCPL &   FCCH1   &   FCCH2   &    NeyProb   &     CLs  \\
\hline
\hline
b  & $ 10 $ & $ 10 $ & $ 0.5 $ & $ 0.5 $ & $ 10 $ & $ 10 $ & $ 10 $  \\
\hline
$\protect\mu $  & $ 20 $ & $ 20 $ & $ 2 $ & $ 1.6 $ & $ 19.6 $ & $ 20 $ & $ 20 $  \\
\hline
Coverage    & $ 93.1 $ & $ 95.9 $ & $ 89.4 $ & $ 86.8 $ & $ 32.5 $ & $ 93.9 $ & $ 91.8 $  \\
\hline
\end{tabular}
\begin{tabular}{|c|c|c|c|c||c|}
\hline
&  $1$ &  $1/\sqrt{\mu }$ & $1/\sqrt{b}$ & $1/\sqrt{\mu b }$ & $1/\sqrt{\mu +b}$ \\
\hline
\hline
b  & $ 10 $ & $ 10 $ & $ 10 $ & $ 10 $ & $ 10 $  \\
\hline
$\protect\mu $  & $ 20 $ & $ 20 $ & $ 20 $ & $ 20 $ & $ 20 $  \\
\hline
Coverage    & $ 91.3 $ & $ 91.9 $ & $ 87.9 $ & $ 89 $ & $ 90.7 $  \\
\hline
\end{tabular}
\end{table}

\begin{figure}[H]
\caption{Coverage and Expected Lengths for  
                  for the case $b=0.5$, $\protect\tau =1$ and $95\%$ confidence intervals}
\label{fig:05-1-95}
\centering
\includegraphics[width=0.98%
\textwidth,natwidth=450,natheight=450]{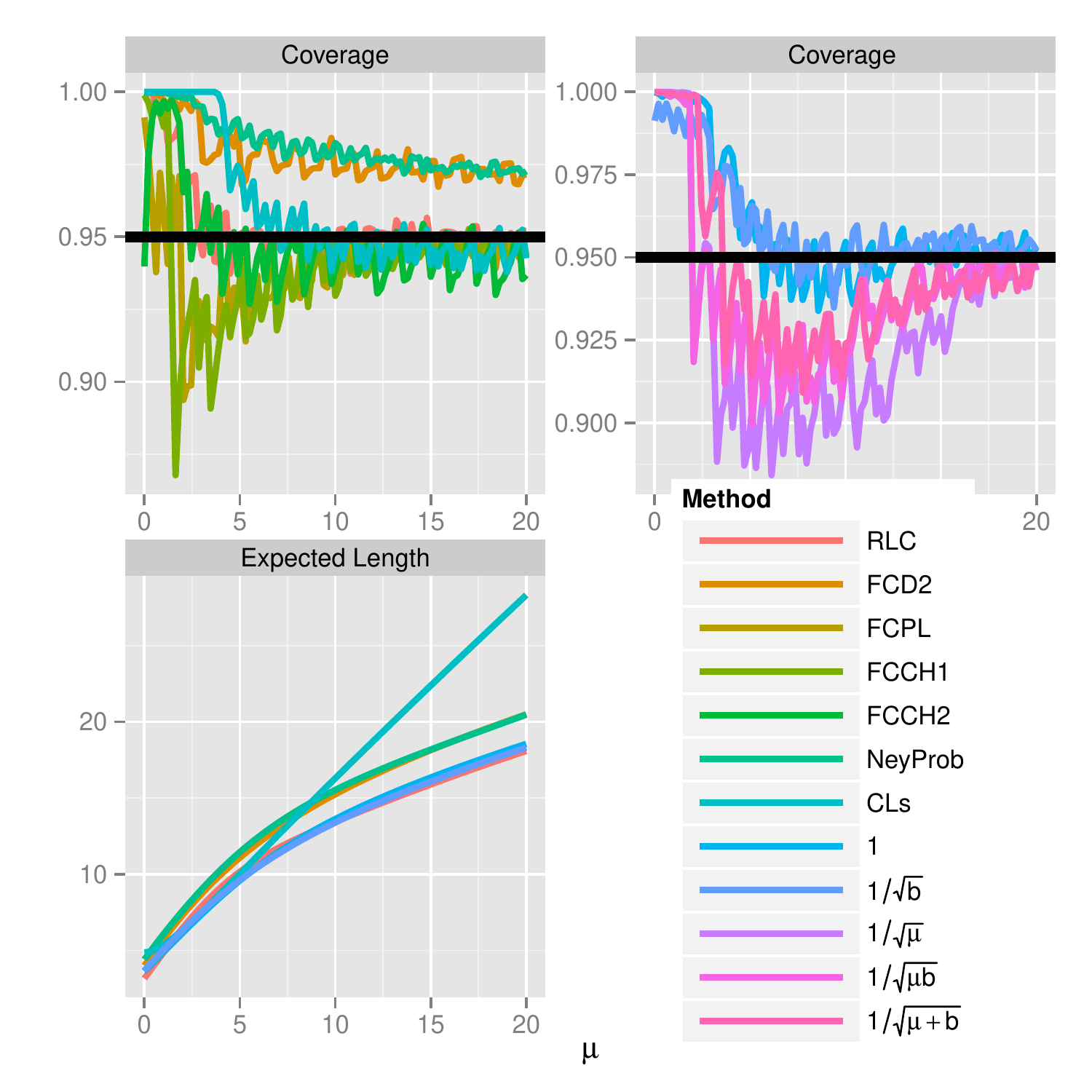}
\end{figure}
\begin{figure}[H]
\caption{Coverage and Expected Lengths for  
                  for the case $b=3$, $\protect\tau =1$ and $95\%$ confidence intervals}
\label{fig:3-1-95}
\centering
\includegraphics[width=0.98%
\textwidth,natwidth=450,natheight=450]{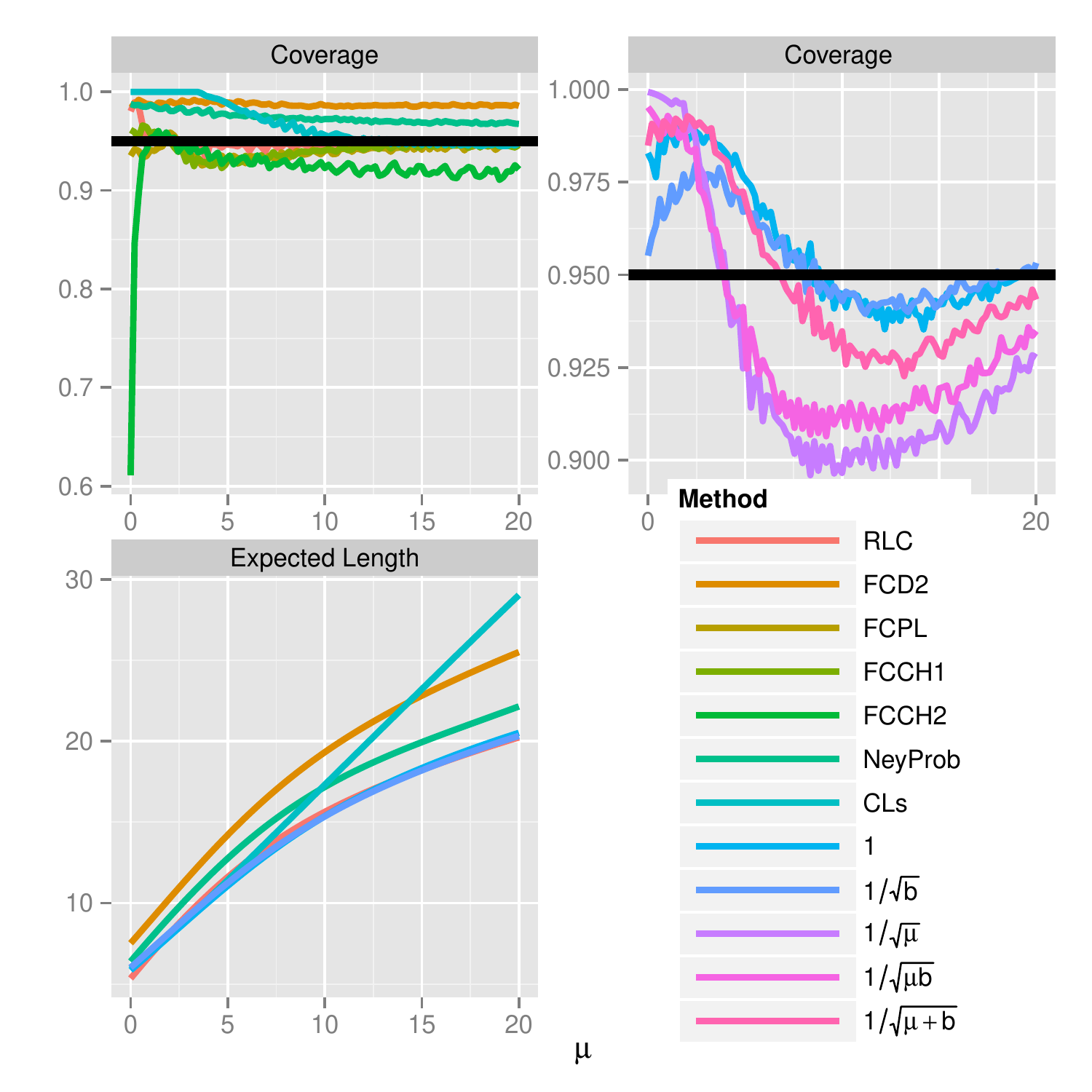}
\end{figure}
\begin{figure}[H]
\caption{Coverage and Expected Lengths for  
                  for the case $b=5$, $\protect\tau =1$ and $95\%$ confidence intervals}
\label{fig:5-1-95}
\centering
\includegraphics[width=0.98%
\textwidth,natwidth=450,natheight=450]{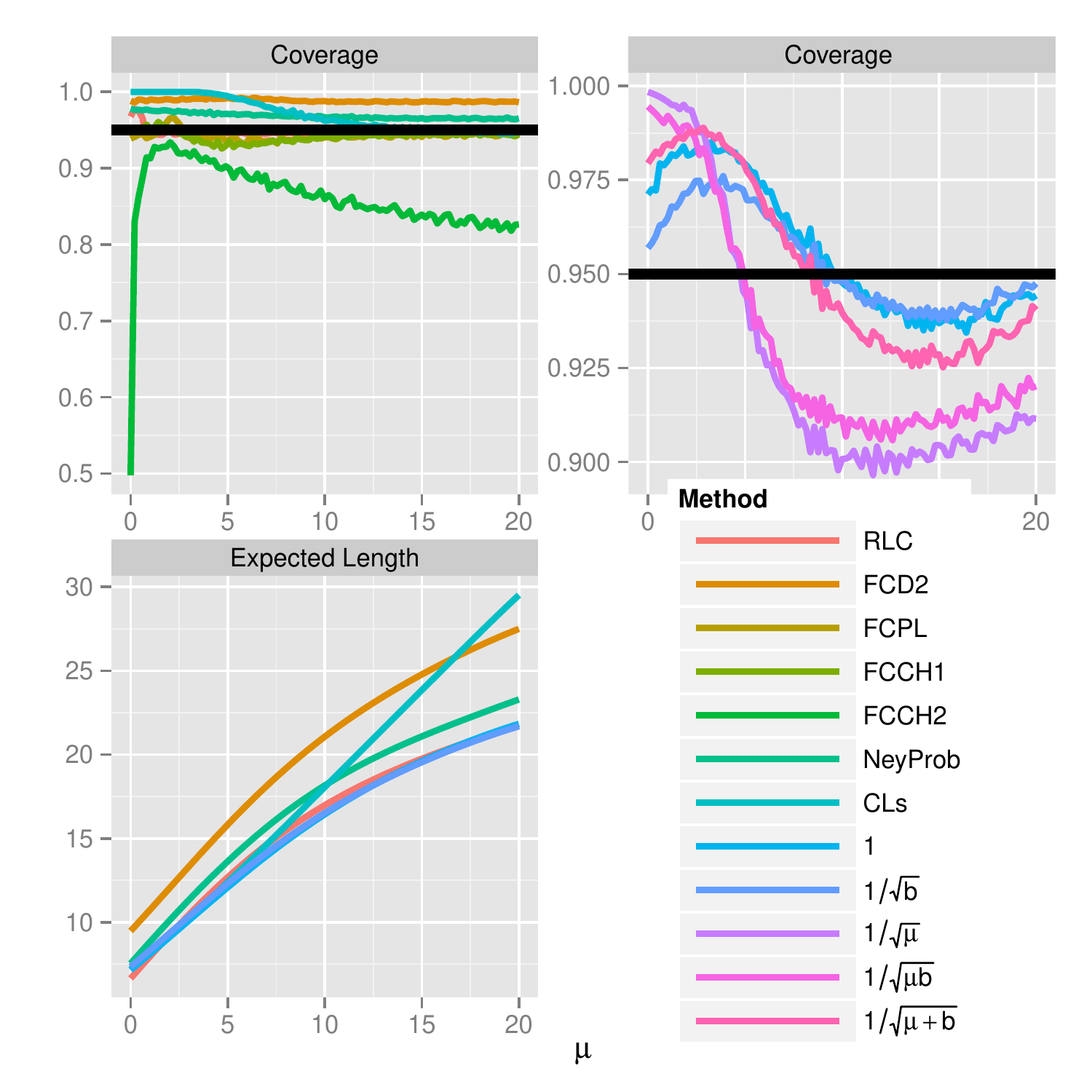}
\end{figure}

\subsubsection{Case $\protect\tau =2 $ and $68\%$ Confidence Intervals}
\begin{table}[tbh]
\caption{Worst coverage of each method, for $0.5\leq b\leq 10$ and $0\leq 
        \protect\mu \leq 20$. $\protect\tau =$ 2 $ and $ 68 $\%$ confidence intervals}
\label{tab:worst2.68}
\begin{tabular}{|c|c|c|c|c|c|c|c|}
\hline
&    RLC   &     FCD2   &   FCPL &   FCCH1   &   FCCH2   &    NeyProb   &     CLs  \\
\hline
\hline
b  & $ 0.5 $ & $ 10 $ & $ 0.5 $ & $ 0.5 $ & $ 10 $ & $ 10 $ & $ 10 $  \\
\hline
$\protect\mu $  & $ 2.4 $ & $ 0 $ & $ 0.8 $ & $ 0.4 $ & $ 20 $ & $ 20 $ & $ 20 $  \\
\hline
Coverage    & $ 62.8 $ & $ 76.3 $ & $ 52.8 $ & $ 47 $ & $ 23.9 $ & $ 67.3 $ & $ 65.5 $  \\
\hline
\end{tabular}
\begin{tabular}{|c|c|c|c|c||c|}
\hline
&  $1$ &  $1/\sqrt{\mu }$ & $1/\sqrt{b}$ & $1/\sqrt{\mu b }$ & $1/\sqrt{\mu +b}$ \\
\hline
\hline
b  & $ 7.9 $ & $ 9.4 $ & $ 0.5 $ & $ 0.5 $ & $ 0.5 $  \\
\hline
$\protect\mu $  & $ 6.1 $ & $ 6.5 $ & $ 0 $ & $ 0 $ & $ 2.9 $  \\
\hline
Coverage    & $ 61.9 $ & $ 62.2 $ & $ 38.8 $ & $ 38.7 $ & $ 56.3 $  \\
\hline
\end{tabular}
\end{table}

\begin{figure}[H]
\caption{Coverage and Expected Lengths for  
                  for the case $b=0.5$, $\protect\tau =2$ and $68\%$ confidence intervals}
\label{fig:05-2-68}
\centering
\includegraphics[width=0.98%
\textwidth,natwidth=450,natheight=450]{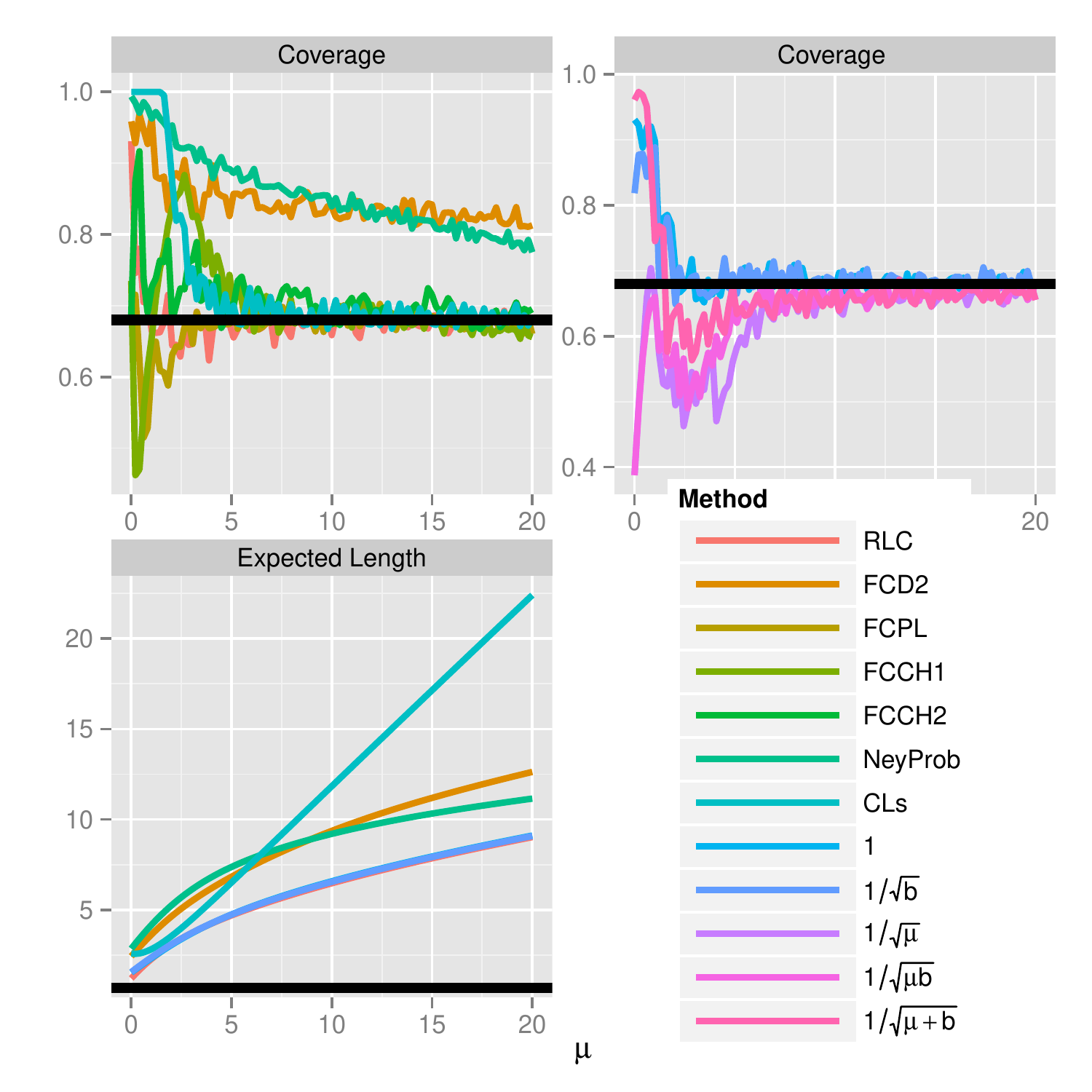}
\end{figure}
\begin{figure}[H]
\caption{Coverage and Expected Lengths for  
                  for the case $b=3$, $\protect\tau =2$ and $68\%$ confidence intervals}
\label{fig:3-2-68}
\centering
\includegraphics[width=0.98%
\textwidth,natwidth=450,natheight=450]{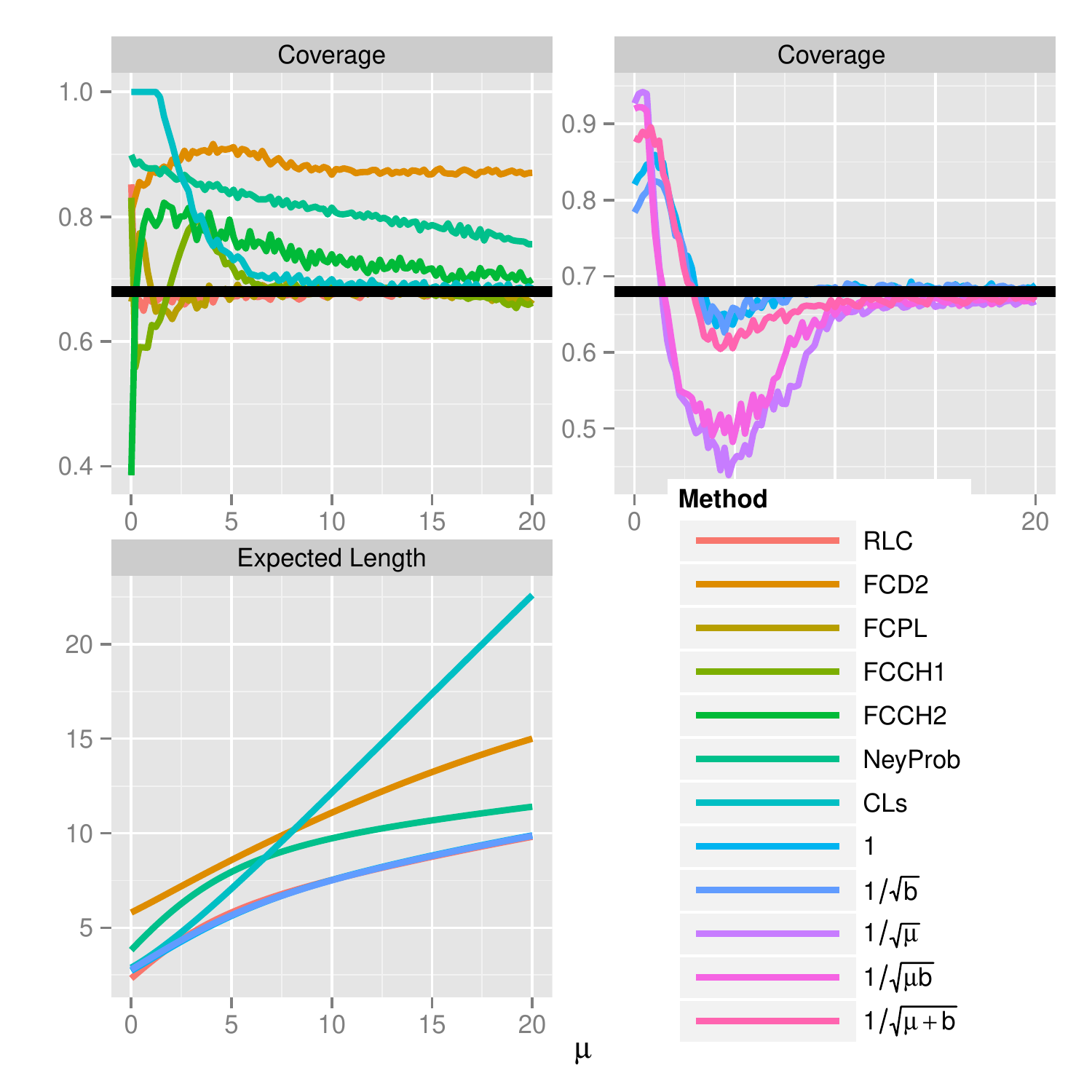}
\end{figure}
\begin{figure}[H]
\caption{Coverage and Expected Lengths for  
                  for the case $b=5$, $\protect\tau =2$ and $68\%$ confidence intervals}
\label{fig:5-2-68}
\centering
\includegraphics[width=0.98%
\textwidth,natwidth=450,natheight=450]{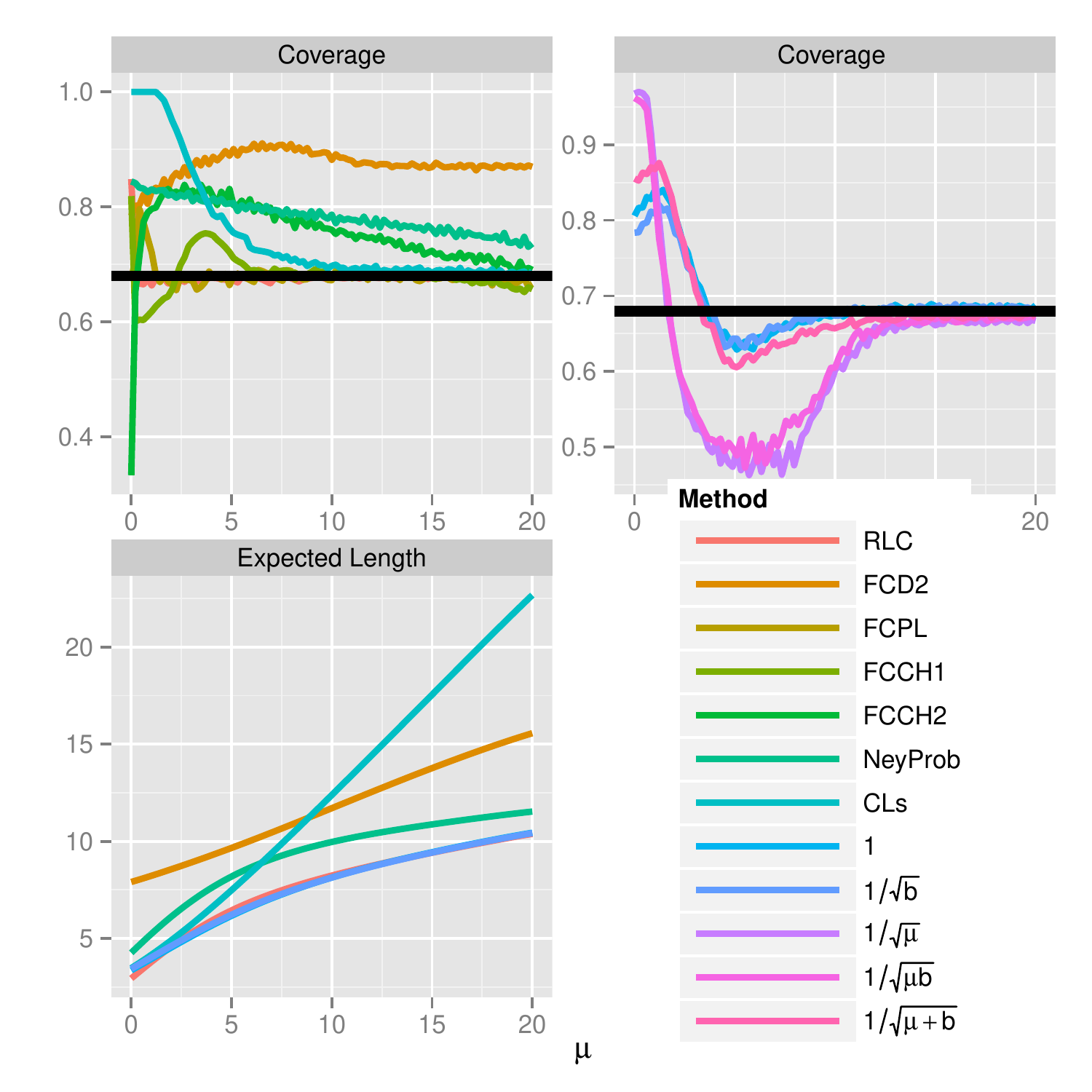}
\end{figure}

\subsubsection{Case $\protect\tau =2 $ and $90\%$ Confidence Intervals}
\begin{table}[tbh]
\caption{Worst coverage of each method, for $0.5\leq b\leq 10$ and $0\leq 
        \protect\mu \leq 20$. $\protect\tau =$ 2 $ and $ 90 $\%$ confidence intervals}
\label{tab:worst2.90}
\begin{tabular}{|c|c|c|c|c|c|c|c|}
\hline
&    RLC   &     FCD2   &   FCPL &   FCCH1   &   FCCH2   &    NeyProb   &     CLs  \\
\hline
\hline
b  & $ 0.5 $ & $ 0.5 $ & $ 0.5 $ & $ 0.5 $ & $ 10 $ & $ 10 $ & $ 10 $  \\
\hline
$\protect\mu $  & $ 2 $ & $ 19.6 $ & $ 1.6 $ & $ 0.4 $ & $ 20 $ & $ 20 $ & $ 20 $  \\
\hline
Coverage    & $ 85.6 $ & $ 93.6 $ & $ 84.8 $ & $ 56.8 $ & $ 30.4 $ & $ 88.1 $ & $ 87.3 $  \\
\hline
\end{tabular}
\begin{tabular}{|c|c|c|c|c||c|}
\hline
&  $1$ &  $1/\sqrt{\mu }$ & $1/\sqrt{b}$ & $1/\sqrt{\mu b }$ & $1/\sqrt{\mu +b}$ \\
\hline
\hline
b  & $ 10 $ & $ 7.7 $ & $ 1.9 $ & $ 9.4 $ & $ 0.5 $  \\
\hline
$\protect\mu $  & $ 13.5 $ & $ 12.2 $ & $ 7.8 $ & $ 11.8 $ & $ 5.3 $  \\
\hline
Coverage    & $ 87.4 $ & $ 87.6 $ & $ 79.7 $ & $ 82.2 $ & $ 84.6 $  \\
\hline
\end{tabular}
\end{table}

\begin{figure}[H]
\caption{Coverage and Expected Lengths for  
                  for the case $b=0.5$, $\protect\tau =2$ and $90\%$ confidence intervals}
\label{fig:05-2-90}
\centering
\includegraphics[width=0.98%
\textwidth,natwidth=450,natheight=450]{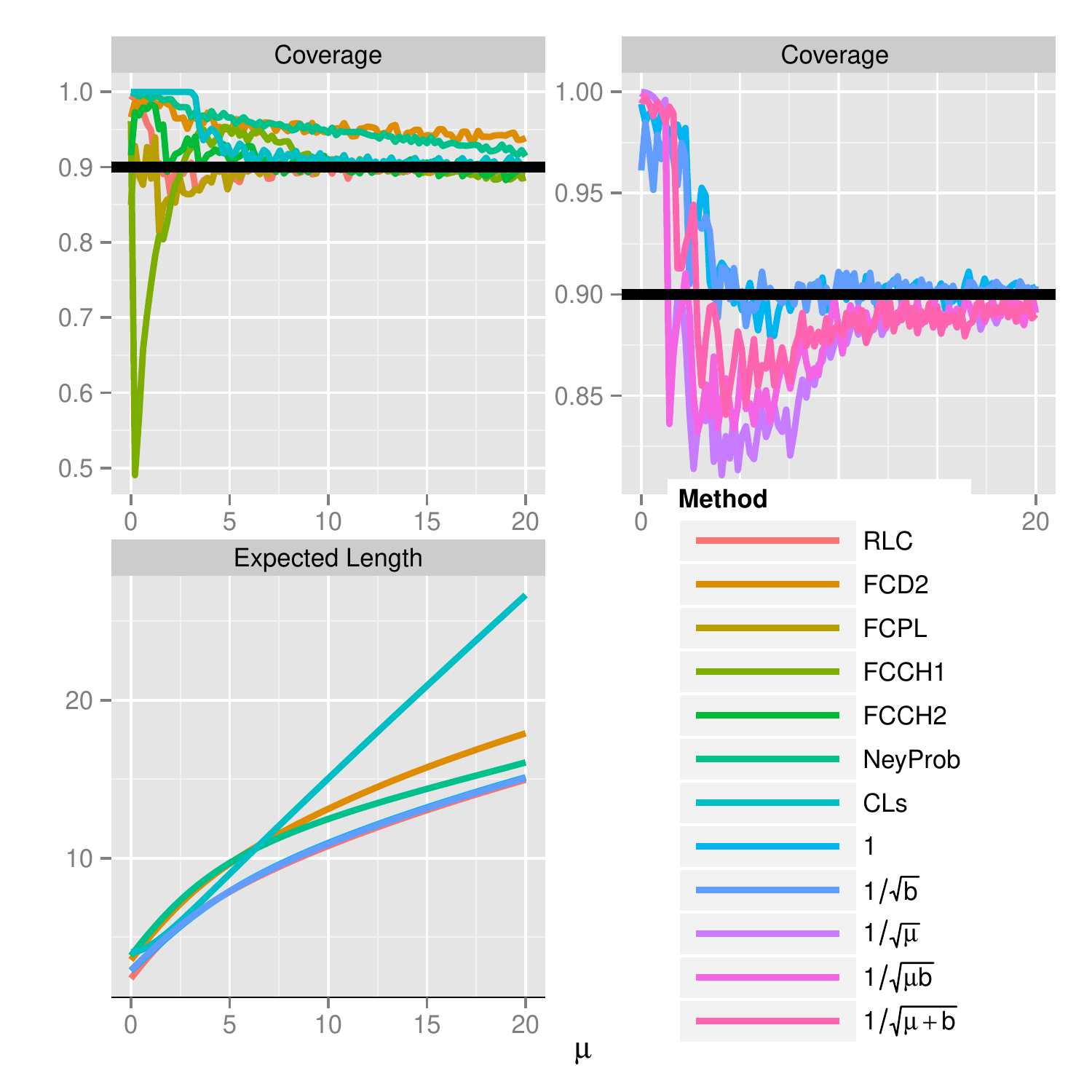}
\end{figure}
\begin{figure}[H]
\caption{Coverage and Expected Lengths for  
                  for the case $b=3$, $\protect\tau =2$ and $90\%$ confidence intervals}
\label{fig:3-2-90}
\centering
\includegraphics[width=0.98%
\textwidth,natwidth=450,natheight=450]{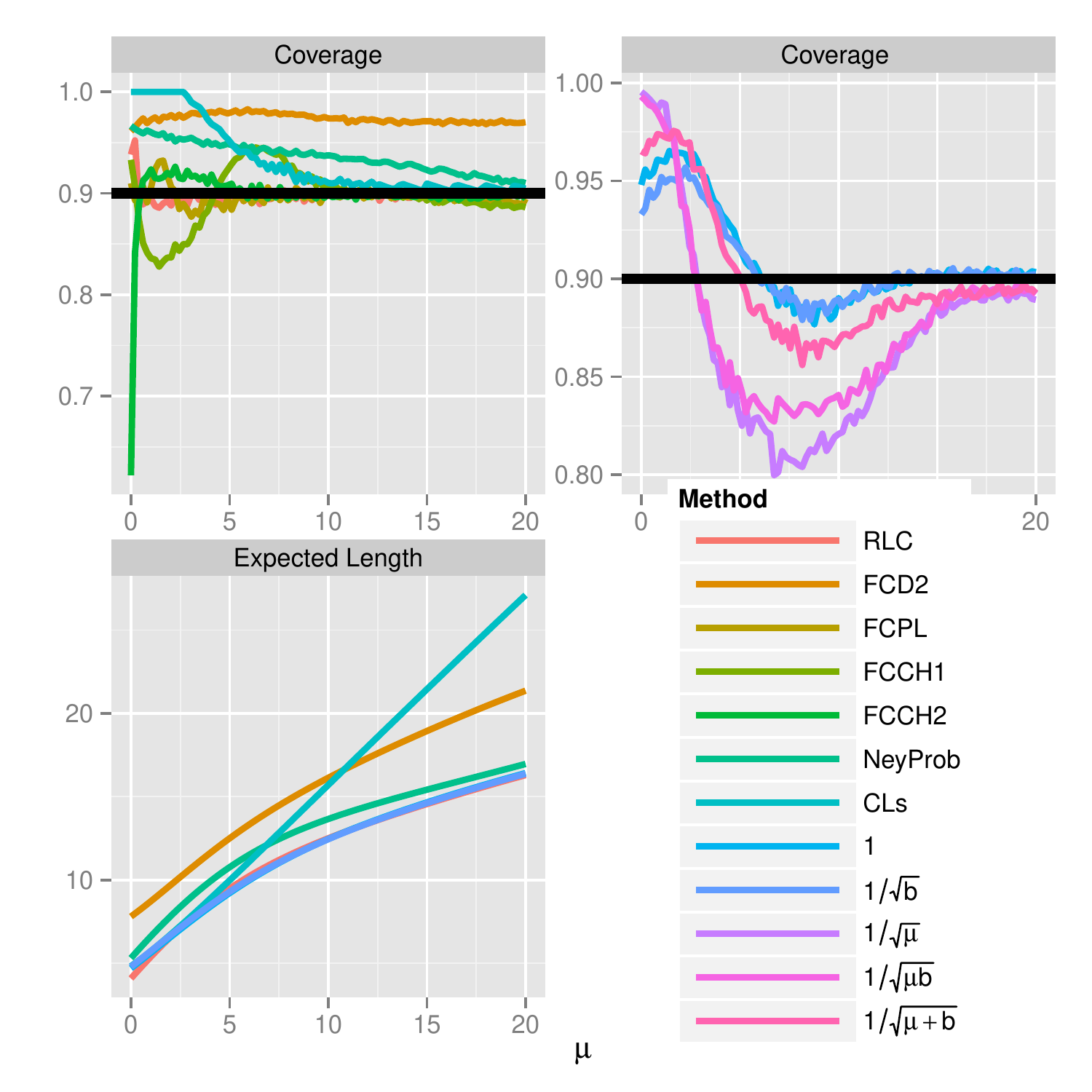}
\end{figure}
\begin{figure}[H]
\caption{Coverage and Expected Lengths for  
                  for the case $b=5$, $\protect\tau =2$ and $90\%$ confidence intervals}
\label{fig:5-2-90}
\centering
\includegraphics[width=0.98%
\textwidth,natwidth=450,natheight=450]{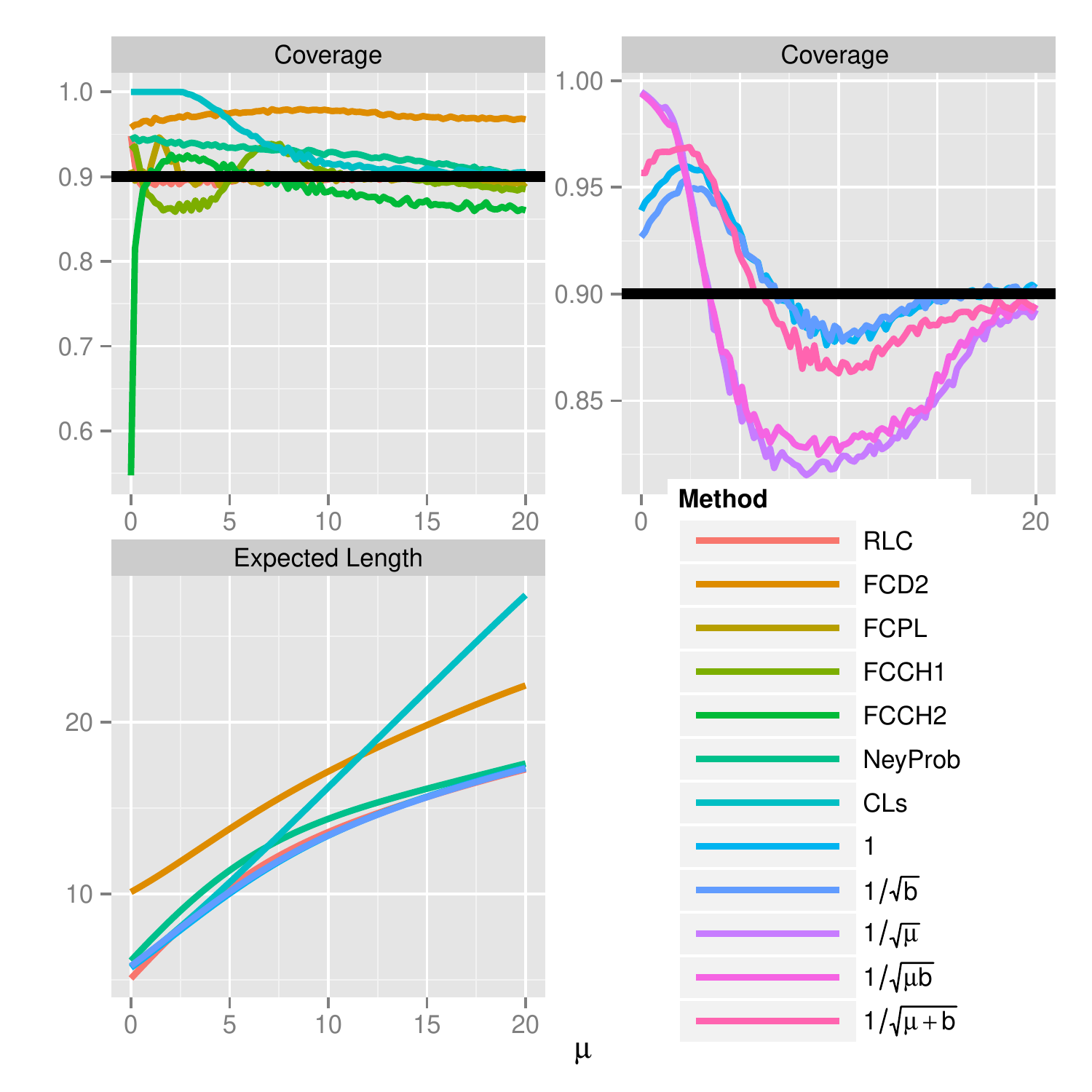}
\end{figure}

\subsubsection{Case $\protect\tau =2 $ and $95\%$ Confidence Intervals}
\begin{table}[tbh]
\caption{Worst coverage of each method, for $0.5\leq b\leq 10$ and $0\leq 
        \protect\mu \leq 20$. $\protect\tau =$ 2 $ and $ 95 $\%$ confidence intervals}
\label{tab:worst2.95}
\begin{tabular}{|c|c|c|c|c|c|c|c|}
\hline
&    RLC   &     FCD2   &   FCPL &   FCCH1   &   FCCH2   &    NeyProb   &     CLs  \\
\hline
\hline
b  & $ 0.5 $ & $ 10 $ & $ 0.5 $ & $ 0.5 $ & $ 10 $ & $ 10 $ & $ 10 $  \\
\hline
$\protect\mu $  & $ 2.9 $ & $ 20 $ & $ 2.4 $ & $ 0.8 $ & $ 20 $ & $ 20 $ & $ 20 $  \\
\hline
Coverage    & $ 93.3 $ & $ 96.5 $ & $ 91.4 $ & $ 72.7 $ & $ 33.8 $ & $ 92.7 $ & $ 92.3 $  \\
\hline
\end{tabular}
\begin{tabular}{|c|c|c|c|c||c|}
\hline
&  $1$ &  $1/\sqrt{\mu }$ & $1/\sqrt{b}$ & $1/\sqrt{\mu b }$ & $1/\sqrt{\mu +b}$ \\
\hline
\hline
b  & $ 10 $ & $ 10 $ & $ 10 $ & $ 1.1 $ & $ 10 $  \\
\hline
$\protect\mu $  & $ 20 $ & $ 20 $ & $ 20 $ & $ 9.4 $ & $ 20 $  \\
\hline
Coverage    & $ 92.5 $ & $ 92.7 $ & $ 89.4 $ & $ 89.6 $ & $ 91.8 $  \\
\hline
\end{tabular}
\end{table}

\begin{figure}[H]
\caption{Coverage and Expected Lengths for  
                  for the case $b=0.5$, $\protect\tau =2$ and $95\%$ confidence intervals}
\label{fig:05-2-95}
\centering
\includegraphics[width=0.98%
\textwidth,natwidth=450,natheight=450]{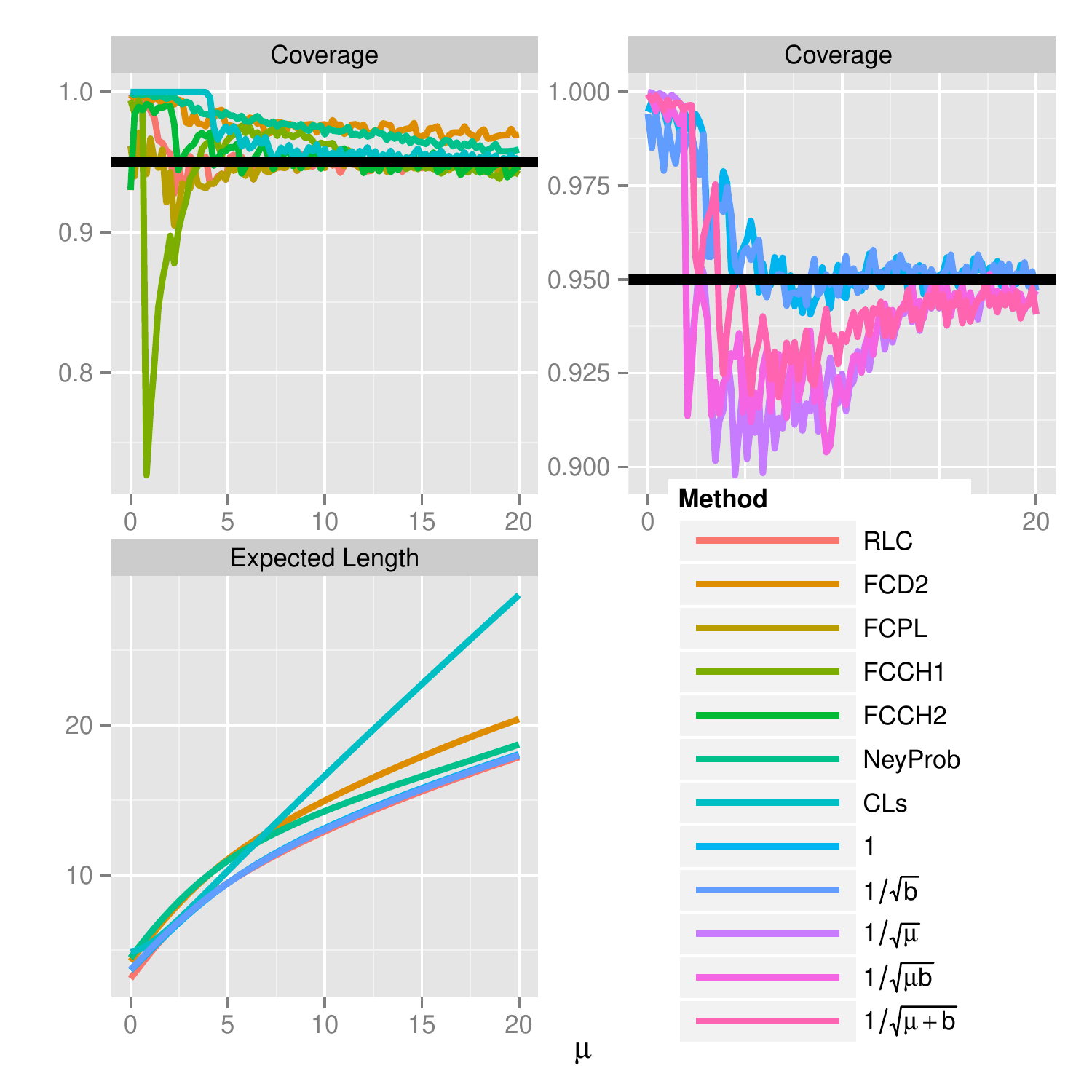}
\end{figure}
\begin{figure}[H]
\caption{Coverage and Expected Lengths for  
                  for the case $b=3$, $\protect\tau =2$ and $95\%$ confidence intervals}
\label{fig:3-2-95}
\centering
\includegraphics[width=0.98%
\textwidth,natwidth=450,natheight=450]{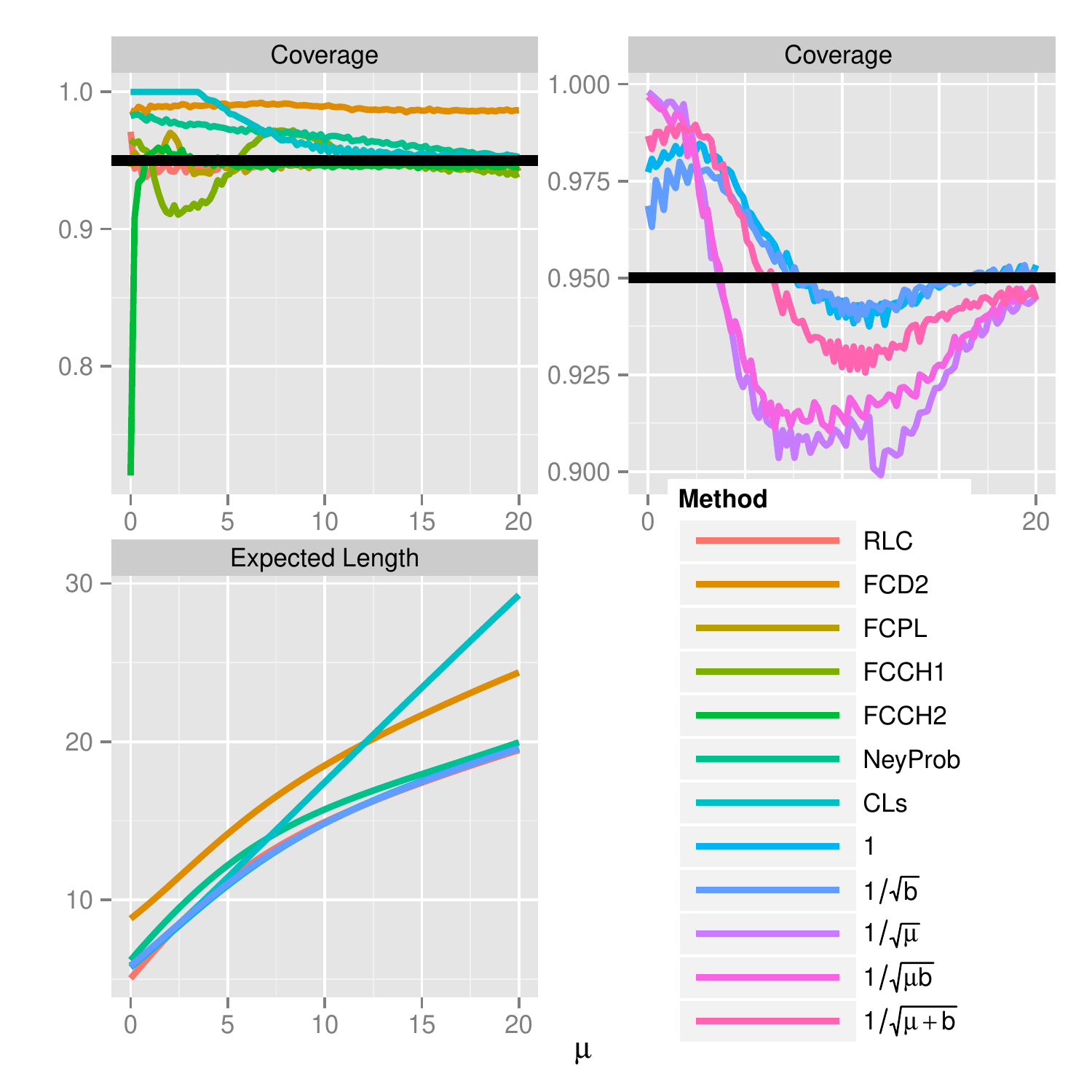}
\end{figure}
\begin{figure}[H]
\caption{Coverage and Expected Lengths for  
                  for the case $b=5$, $\protect\tau =2$ and $95\%$ confidence intervals}
\label{fig:5-2-95}
\centering
\includegraphics[width=0.98%
\textwidth,natwidth=450,natheight=450]{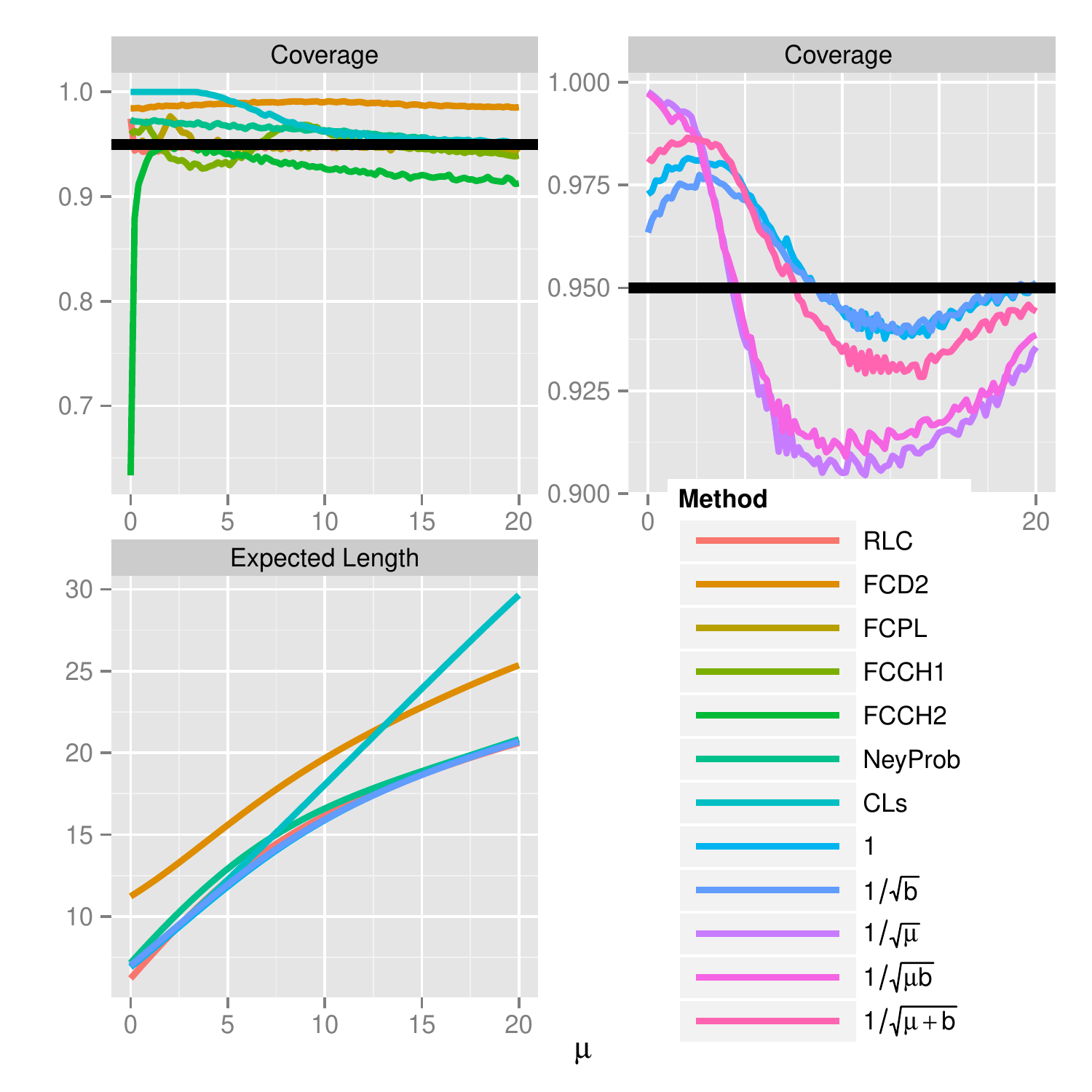}
\end{figure}

\end{document}